\documentclass[english,prb,notitlepage,nofootinbib,twocolumn,superscriptaddress]{revtex4-1}
\usepackage[T1]{fontenc}
\usepackage[latin9]{inputenc}
\usepackage{geometry}
\usepackage{color}
\usepackage{babel}
\usepackage{float}
\usepackage{amsmath}
\usepackage{graphicx}
\usepackage[colorlinks=true,breaklinks]{hyperref}

\makeatletter
\newcommand{\lyxdot}{.}
\makeatother

\begin{document}

\title{The Fierz convergence criterion: a controlled approach to strongly-interacting
systems with small embedded clusters }

\author{Thomas Ayral}

\affiliation{Physics and Astronomy Department, Rutgers University, Piscataway,
NJ 08854, USA}

\affiliation{Institut de Physique Théorique (IPhT), CEA, CNRS, UMR 3681, 91191
Gif-sur-Yvette, France}

\author{Jaksa Vu\v{c}i\v{c}evi\'{c}}

\affiliation{Institut de Physique Théorique (IPhT), CEA, CNRS, UMR 3681, 91191
Gif-sur-Yvette, France}

\affiliation{Scientific Computing Laboratory, Center for the Study of Complex
Systems, Institute of Physics Belgrade, University of Belgrade, Pregrevica
118, 11080 Belgrade, Serbia}

\author{Olivier Parcollet}

\affiliation{Institut de Physique Théorique (IPhT), CEA, CNRS, UMR 3681, 91191
Gif-sur-Yvette, France}
\begin{abstract}
We present an embedded-cluster method, based on the TRILEX formalism. 
It turns the Fierz ambiguity, inherent to approaches based on a
bosonic decoupling of local fermionic interactions, into a convergence
criterion. It is based on the approximation of the three-leg vertex
by a coarse-grained vertex computed by solving a self-consistently
determined multi-site effective impurity model. The computed self-energies
are, by construction, continuous functions of momentum. We show that,
in three interaction and doping regimes of parameters of the two-dimensional
Hubbard model, self-energies obtained with clusters of size four only are very close to numerically exact
benchmark results. We show that the
Fierz parameter, which parametrizes the freedom in the Hubbard-Stratonovich
decoupling, can be used as a quality control parameter. By contrast,
the $GW$+extended dynamical mean field theory approximation with
four cluster sites is shown to yield good results only in the weak-coupling
regime and for a particular decoupling. Finally, we show that the
vertex has spatially nonlocal components only at low Matsubara frequencies. 
\end{abstract}
\maketitle
Two major approaches have been put forth to fathom the nature of high-temperature
superconductivity. Spin fluctuation theory\cite{Chubukov2002,Onufrieva2009,Metlitski2010,Efetov2013,Onufrieva2012,Scalapino2012,Wang2014,Wang2016},
inspired by the early experiments on cuprate compounds, is based on
the introduction of phenomenological bosonic fluctuations coupled
to the electrons. It belongs to a larger class of methods, including
the fluctuation-exchange (FLEX)\cite{Bickers1989} and $GW$ approximations\cite{Hedin1965,Hedin1999},
or the Eliashberg theory of superconductivity\cite{Eliashberg1960}.
In the Hubbard model, these methods can formally be obtained by decoupling
the electronic interactions with Hubbard-Stratonovich (HS) bosons carrying
charge, spin or pairing fluctuations. They are particularly well suited
for describing the system's long-range modes. However, they suffer
from two main drawbacks: without an analog of Migdal's theorem for
spin fluctuations, they are quantitatively uncontrolled; worse, the
results depend on the precise form of the bosonic fluctuations used
to decouple the interaction term, an issue referred to as the ``Fierz
ambiguity''\cite{Jaeckel2003,Baier2004,Bartosch2009,Borejsza2003,Borejsza2004,Dupuis2002}.

A second class of methods, following Anderson\cite{Anderson1987},
puts primary emphasis on the fact that the undoped compounds are Mott
insulators, where local physics plays a central role. Approaches like
dynamical mean field theory (DMFT)\cite{Georges1996} and its cluster
extensions\cite{Hettler1998,Hettler1999,Lichtenstein2000,Kotliar2001,Maier2005a},
which self-consistently map the lattice problem onto an effective
problem describing a cluster of interacting atoms embedded in a noninteracting
host, are tools of choice to examine Anderson's idea. Cluster DMFT
has indeed been shown to give a consistent qualitative picture of
cuprate physics, including pseudogap and superconducting phases\cite{Kyung2009,Sordi2012,Civelli2008,Ferrero2010,Gull2013,Macridin2004,Maier2004,Maier2005,Maier2006,Gull2010,Yang2011,Macridin2008,Macridin2006,Jarrell2001,Parcollet2004,Werner2009,Biroli2004,Bergeron2011,Kyung2004,Kyung2006a,Okamoto2010,Sordi2010,Sordi2012a,Civelli2005,Ferrero2008,Ferrero2009,Gull2009,Chen2015,Chen2016}.
Compared to fluctuation theories, it \emph{a priori} comes with a
control parameter, the size $N_{c}$ of the embedded cluster. However,
this is of limited practical use, since the convergence with $N_{c}$
is nonmonotonic for small $N_{c}$\cite{Maier2005}, requiring large
$N_{c}$'s, which cannot be reached in interesting physical regimes
due to the Monte-Carlo negative sign problem. Thus, converged cluster
DMFT results can only be obtained at high temperatures\cite{Leblanc2015}.
There, detailed studies\cite{Gunnarsson,Gunnarsson2016,Wu2016} point
to the importance of (possibly long-ranged) spin fluctuations, calling
for a unification of both classes of approaches. First steps in this
direction have been accomplished by diagrammatic extensions of DMFT\cite{Rohringer2017,Biermann2003,Sun2002,Sun2004,Ayral2012,Ayral2013,Biermann2014,Ayral2017,Rubtsov2008,Rubtsov2011,VanLoon2014,Stepanov2015,Toschi2007,Katanin2009,Schafer2014,Valli2014,Li2015a,Rohringer2016,Ayral2016,Hafermann2008,Slezak2009,Yang2011c},
and by the single-site TRILEX formalism\cite{Ayral2015,Ayral2015c},
which interpolates between long-range and Mott physics, and describes
aspects of pseudogap physics and the $d$-wave superconducting dome\cite{Vucicevic2017}.

In this Letter, we turn the Fierz ambiguity into a convergence criterion
in the cluster extension of TRILEX. Like fluctuation approaches,
cluster TRILEX is based on the introduction of bosonic degrees of freedom. Like
cluster DMFT, it maps the corresponding electron-boson
problem onto a cluster impurity problem. The latter is solved for
its three-leg vertex, which is used as a cluster vertex correction
to the self-energies. This approach improves on fluctuation
approaches by endowing them with a control parameter, thus curing
the absence of a Migdal theorem. In some parameter regimes, it can
solve the cluster DMFT large-$N_{c}$ stalemate by instead requiring
minimal sensitivity to the Fierz parameter as a convergence criterion
of the solution.

To illustrate the method, we focus on the two-dimensional Hubbard
model, the simplest model to describe high-temperature superconductors.
It is defined by the Hamiltonian: 
\begin{equation}
H=\sum_{ij\sigma}t_{ij}c_{i\sigma}^{\dagger}c_{j\sigma}+U\sum_{i}n_{i\uparrow}n_{i\downarrow}\label{eq:Hubbard_model}
\end{equation}

where $c_{i\sigma}^{\dagger}$ ($c_{i\sigma}$) creates (annihilates)
an electron of spin $\sigma$ at Bravais site $\mathbf{r}_{i}$,
$t_{ij}$ is the hopping matrix (with {[}next-{]}nearest-neighbor
hopping parametrized by $t$ {[}$t'${]}), and $U$ the local electronic repulsion. We set $t=-0.25$ and use $D\equiv 4|t|$ as the energy unit.

The first step of the TRILEX method consists in decoupling the interaction
term with HS fields. There are several possible such decouplings, a fact called the Fierz ambiguity. Here,
we choose
\footnote{See Suppl. Mat. A for another choice}
to express the interaction in the charge and longitudinal
spin channel (``Ising decoupling''), i.e, up to a density term:
\begin{equation}
Un_{i\uparrow}n_{i\downarrow}=\frac{1}{2}U^{\mathrm{ch}}n_{i}n_{i}+\frac{1}{2}U^{\mathrm{sp}}s_{i}^{z}s_{i}^{z}\label{eq:U_decoupling}
\end{equation}

with $n\equiv n_{\uparrow}+n_{\downarrow}$ and $s^{z}\equiv n_{\uparrow}-n_{\downarrow}$.This
holds provided $U^{\mathrm{ch}}-U^{\mathrm{sp}}=U$, or equivalently
\begin{align}
U^{\mathrm{ch}} & =\alpha U,\;\;U^{\mathrm{sp}}=(\alpha-1)U\label{eq:U_Fierz-ising}
\end{align}

The ``Fierz parameter'' $\alpha$ materializes the freedom in choosing
the charge-to-spin fluctuation ratio. The right-hand side
of Eq. (\ref{eq:U_decoupling}) is decoupled with a charge and a spin boson,
 resulting in an electron-boson coupling problem\cite{Ayral2015,Ayral2015c}. Its fermionic and bosonic interacting
Green's functions are given by Dyson equations:\begin{subequations}
\begin{align}
G(\mathbf{k},i\omega) & =\frac{1}{i\omega+\mu-\varepsilon(\mathbf{k})-\Sigma(\mathbf{k},i\omega)}\label{eq:Dyson_G_lattice}\\
W^{\eta}(\mathbf{q},i\Omega) & =\frac{U^{\eta}}{1-U^{\eta}P^{\eta}(\mathbf{q},i\Omega)}\label{eq:Dyson_W_lattice}
\end{align}
\end{subequations}$\varepsilon(\mathbf{k})$ is the Fourier transform
of $t_{ij}$ ($\varepsilon(\mathbf{k})=2t\left(\cos(k_{x})+\cos(k_{y})\right)+4t'\cos(k_{x})\cos(k_{y})$),
$\mu$ the chemical potential, $\eta=\mathrm{ch},\mathrm{sp}$, and
$i\omega$ {[}resp. $i\Omega${]} denote fermionic {[}resp. bosonic{]}
Matsubara frequencies. The self-energy $\Sigma(\mathbf{k},i\omega)$
and polarization $P^{\eta}(\mathbf{q},i\Omega)$ are given by the
exact Hedin expressions:\begin{subequations}

\begin{align}
 &\Sigma(\mathbf{k},i\omega)  =\label{eq:Sigma_exact}\\
 &\;\; -\sum_{\eta}\sum_{\mathbf{q},i\Omega}G(\mathbf{k}+\mathbf{q},i\omega+i\Omega)W^{\eta}(\mathbf{q},i\Omega)\Lambda_{\mathbf{k}\mathbf{q}}^{\eta}(i\omega,i\Omega)\nonumber\\
& P^{\eta}(\mathbf{q},i\Omega)  =\label{eq:P_exact}\\
 &\;\; 2\sum_{\mathbf{k},i\omega}G(\mathbf{k}+\mathbf{q},i\omega+i\Omega)G(\mathbf{k},i\omega)\Lambda_{\mathbf{k}\mathbf{q}}^{\eta}(i\omega,i\Omega)\nonumber
\end{align}

\end{subequations} $\Lambda_{\mathbf{k}\mathbf{q}}^{\eta}(i\omega,i\Omega)$ is the interacting electron-boson vertex. TRILEX approximates
it with a vertex computed from a self-consistent impurity model. In
previous works\cite{Ayral2015,Ayral2015c}, this impurity model contained
a single site.

There are several ways to extend the TRILEX method to cluster impurity
problems, like in DMFT. Here, we consider the analog of the dynamical
cluster approximation (DCA\cite{Hettler1998,Hettler1999,Maier2005a}), and use periodic clusters so as not to break the lattice translational
symmetry, at the price of discontinuities
in the momentum dependence of the \emph{vertex function}. Other cluster
variants such as a real-space version, inspired from cellular DMFT\cite{Lichtenstein2000,Kotliar2001},
are also possible, but break translation invariance and require arbitrary
reperiodization procedures.

\begin{figure}
\begin{centering}
\includegraphics[width=0.85\columnwidth]{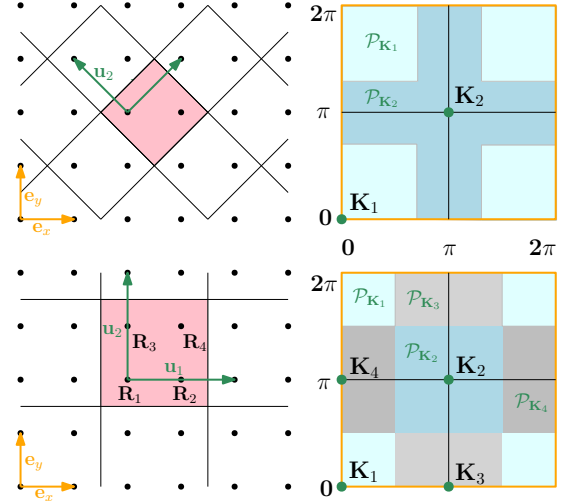} 
\par\end{centering}
\caption{Cluster geometry: real (left) and reciprocal (right) space, for $N_{c}=2$
(top) and $N_{c}=4$ (bottom). $\mathbf{e}_{x}$ and $\mathbf{e}_{y}$
($\mathbf{u}_{1}$ and $\mathbf{u}_{2}$) are the unit vectors of
the Bravais (super)lattice. The colored patches $\mathcal{P}_{\mathbf{K}_{i}}$
are of equal area\label{fig:Cluster-geometry}}
\end{figure}

We straighforwardly generalize the single-site impurity model of TRILEX
to a cluster impurity model defined by the action:

\begin{align}
& S_{\mathrm{imp}} \equiv\iint_{\tau\tau'}\sum_{ij\sigma}c_{i\sigma\tau}^{*}\left\{ -\left[\mathcal{G}^{-1}\right]_{ij}(\tau-\tau')\right\} c_{j\sigma\tau'}\label{eq:S_imp_def}\\
 & +\frac{1}{2}\iint_{\tau\tau'}\sum_{ij} \left\{ n_{i\tau} \mathcal{U}_{ij}^{\mathrm{ch}}(\tau-\tau') n_{j\tau'}+ s_{i\tau}^{z}\mathcal{U}_{ij}^{\mathrm{sp}}(\tau-\tau') s_{j\tau'}^{z}\right \} \nonumber
\end{align}

The latin indices $i,j=1\dots N_{c}$ stand for the cluster positions
$\mathbf{R}_{i}$, $\mathbf{R}_{j}$ (shown in Fig. \ref{fig:Cluster-geometry}
along with the cluster momenta $\{\mathbf{K}_{i}\}_{i=1\dots N_{c}}$).
$c_{i\sigma\tau}^{*}$ and $c_{i\sigma\tau}$ are conjugate Grassmann
fields, $\tau$ denotes imaginary time. Since we have introduced
a charge and a spin bosonic mode,
the impurity action contains interactions in both channels ($\mathcal{U}^{\mathrm{ch}}(\tau)$ and $\mathcal{U}^{\mathrm{sp}}(\tau)$). They
are\emph{ a priori} retarded due to the nonlocal character of $P^{\eta}(\mathbf{q},i\Omega)$.

This cluster impurity model is used to compute the cluster impurity
vertex $\Lambda_{\mathrm{imp}}^{\eta}(\mathbf{K},\mathbf{Q};i\omega,i\Omega)$
with a continuous-time quantum Monte-Carlo algorithm with a hybridization [resp. interaction] expansion for $N_c=1$ [resp. $N_c=2,4$]
(as described in Suppl. Mat. B.3). Next, in the spirit
of DCA, we want to use $\Lambda_{\mathrm{imp}}^{\eta}(\mathbf{K},\mathbf{Q};i\omega,i\Omega)$
to approximate the momentum dependence of the lattice vertex $\Lambda_{\mathbf{k}\mathbf{q}}^{\eta}(i\omega,i\Omega)$
by a coarse-graining procedure. We recall that DCA consists
in coarse-graining the cluster \emph{self-energy }as $\Sigma(\mathbf{k},i\omega)\approx\sum_{\mathbf{K}}\theta_{\mathbf{K}}(\mathbf{k})\Sigma_{\mathrm{imp}}(\mathbf{K},i\omega)$,
where $\Sigma_{\mathrm{imp}}(\mathbf{K},i\omega)$ is the cluster
impurity self-energy, and $\theta_{\mathbf{K}}(\mathbf{k})=1$ if
$\mathbf{k}$ belongs to Brillouin-zone patch $\mathcal{P}_{\mathbf{K}}$, and vanishes
otherwise. For the vertex function, the passage from $\Lambda_{\mathrm{imp}}^{\eta}(\mathbf{K},\mathbf{Q};i\omega,i\Omega)$
to an approximate lattice vertex $\Lambda_{\mathbf{k},\mathbf{q}}^{\eta}(i\omega,i\Omega)$
is not as straightforward. There are several possible coarse-grainings
for the vertex that reduce to single-site TRILEX for $N_c=1$ and are exact in the  $N_c=\infty$ limit, \emph{e.g.}\begin{subequations} 
\begin{align}
\Lambda_{\mathbf{k}\mathbf{q}}^{\eta}(i\omega,i\Omega) & \approx\sum_{\mathbf{K},\mathbf{Q}}\theta_{\mathbf{K}+\mathbf{Q}}(\mathbf{k}+\mathbf{q})\theta_{\mathbf{Q}}(\mathbf{q})\Lambda_{\mathrm{imp}}^{\eta}(\mathbf{K},\mathbf{Q};i\omega,i\Omega)\label{eq:Lambda_periodization_Sigma}\\
\Lambda_{\mathbf{k}\mathbf{q}}^{\eta}(i\omega,i\Omega) & \approx\sum_{\mathbf{K},\mathbf{Q}}\theta_{\mathbf{K}}(\mathbf{k})\theta_{\mathbf{K}+\mathbf{Q}}(\mathbf{k}+\mathbf{q})\Lambda_{\mathrm{imp}}^{\eta}(\mathbf{K},\mathbf{Q};i\omega,i\Omega)\label{eq:Lambda_periodization_P}
\end{align}
\end{subequations}We use a different coarse-graining for $\Sigma$ and for $P$: we substitute (\ref{eq:Lambda_periodization_Sigma})
in (\ref{eq:Sigma_exact}) [resp. (\ref{eq:Lambda_periodization_P}) in (\ref{eq:P_exact})] to compute $\Sigma(\mathbf{k},i\omega)$ [resp. $P^{\eta}(\mathbf{q},i\Omega)$], whence:\begin{subequations}
\begin{align}
& \Sigma(\mathbf{k},i\omega) =\label{eq:Sigma_k_DCA_full}\\
  &-\sum_{\eta,\mathbf{K},\mathbf{Q}}\sum_{\mathbf{q},i\Omega}G_{\mathbf{k}+\mathbf{q}}^{\mathbf{K}+\mathbf{Q}}(i\omega+i\Omega) W_{\mathbf{q}}^{\eta,\mathbf{Q}}(i\Omega)\Lambda_{\mathrm{imp}}^{\eta}(\mathbf{K},\mathbf{Q};i\omega,i\Omega)\nonumber\\
& P^{\eta}(\mathbf{q},i\Omega) =\label{eq:P_Q_DCA_full}\\
 & 2\sum_{\mathbf{K},\mathbf{Q}}\sum_{\mathbf{k},i\omega}G_{\mathbf{k}+\mathbf{q}}^{\mathbf{K}+\mathbf{Q}}(i\omega+i\Omega)G_{\mathbf{k}}^{\mathbf{K}}(i\omega)\Lambda_{\mathrm{imp}}^{\eta}(\mathbf{K},\mathbf{Q};i\omega,i\Omega)\nonumber
\end{align}
\end{subequations}with $X_{\mathbf{k}}^{\mathbf{K}}(i\omega)\equiv\theta_{\mathbf{K}}(\mathbf{k})X(\mathbf{k},i\omega)$ (for $X=G$ and $W$).
As convolutions of continuous functions of $\mathbf{k}$ ($G$ and
$W$) with a piecewise-constant function ($\Lambda$), $\Sigma$ and
$P$ are continuous in $\mathbf{k}$ by construction.

Finally, the cluster dynamical mean fields $\mathcal{G}_{ij}(\tau)$
and $\mathcal{U}_{ij}^{\eta}(\tau)$ are determined by imposing 
the following self-consistency conditions:\begin{subequations}
\begin{align}
G_{\mathrm{imp}}(\mathbf{K},i\omega)[\mathcal{G},\mathcal{U}] & =G_{\mathbf{K}}(i\omega)\label{eq:sc_G}\\
W_{\mathrm{imp}}^{\eta}(\mathbf{Q},i\Omega)[\mathcal{G},\mathcal{U}] & =W_{\mathbf{Q}}^{\eta}(i\Omega)\label{eq:sc_W}
\end{align}
\end{subequations}The left-hand sides are computed by solving the
impurity model. The right-hand sides are the patch-averaged lattice
Green's functions: \begin{subequations} 
\begin{align}
G_{\mathbf{K}}(i\omega) & \equiv\sum_{\mathbf{k}\in\mathcal{P}_{\mathbf{K}}}G(\mathbf{k},i\omega)\label{eq:G_K_def}\\
W_{\mathbf{Q}}^{\eta}(i\Omega) & \equiv\sum_{\mathbf{q}\in\mathcal{P}_{\mathbf{Q}}}W^{\eta}(\mathbf{q},i\Omega)\label{eq:W_Q_def}
\end{align}
\end{subequations}The determination of $\mathcal{G}$ and $\mathcal{U}^{\eta}$
satisfying Eq. (\ref{eq:sc_G}-\ref{eq:sc_W}) is done by forward
recursion (see Suppl. Mat. B.2).

We have implemented this method and studied it in three physically
distinct parameter regimes: (A) \emph{ Weak-coupling regime (}$U/D=0.5$,
$\delta=0\%$, $\beta D=16$, $t'=0$\emph{) at half-filling, }(B)
\emph{Intermediate-coupling regime (}$U/D=1$, $\delta=20\%$, $\beta D=16$,
$t'=0$\emph{) at large doping, }(C) \emph{Strong-coupling regime
(}$U/D=1.4$, $\delta=4\%$, $\beta D=8$, $t'/t=-0.3$\emph{) at
small doping} (the Mott transition occurs at $U_{c}/D\approx1.5$
within plaquette cellular DMFT\cite{Park2008}).
We solve at point A, B, C for different values of $\alpha$.

In the absence of any approximation, every HS decoupling, hence every value of $\alpha$,
yields the same result: the exact solution does not depend on $\alpha$.
The cluster TRILEX approximation  \emph{a priori} breaks
this property, but as $N_{c}$ increases, we expect the  $\alpha$-dependence to become weaker.
We propose to use the weak $\alpha$-dependence for a given $N_c$, \emph{i.e.}
the existence of a plateau for at least a range $\alpha$,
as a (Fierz) convergence criterion.
Whether this criterion is \emph{sufficient} to establish convergence
is an assumption, which we test here using exact benchmarks for points A, B and C.
Indeed, at these temperatures, interactions and dopings, determinant
quantum Monte Carlo (QMC) and/or DCA can be converged and give a numerically
exact solution of the Hubbard
model, albeit at a significant numerical cost. 

\begin{figure}
\begin{centering}
\includegraphics[width=1\columnwidth]{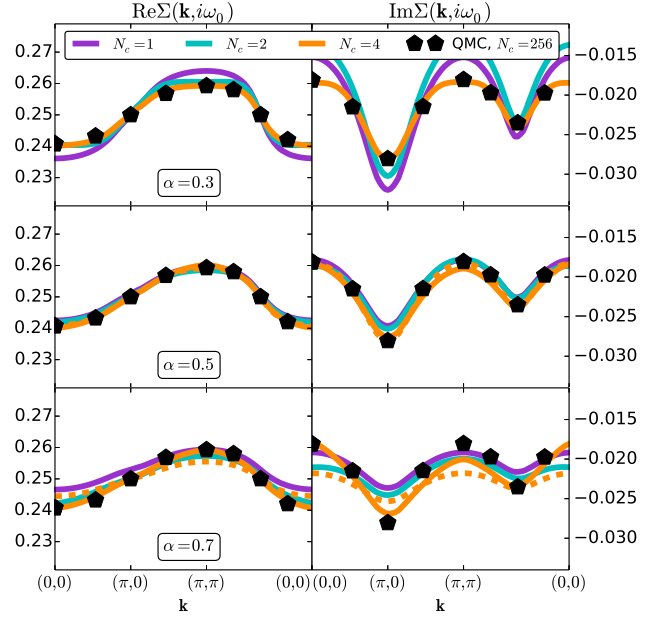} 
\par\end{centering}
\caption{Point A ($U/D=0.5$, $\delta=0\%$, $\beta D=16$, $t'=0$). $\mathrm{Re}\Sigma(\mathbf{k},i\omega_{0})$
(left) and $\mathrm{Im}\Sigma(\mathbf{k},i\omega_{0})$ (right) for $N_{c}=1,2,4$ for various values of $\alpha$ (from top
to bottom), along the path $(0,0)-(\pi,0)-(\pi,\pi)-(0,0)$. Solid
lines: TRILEX. Dashed lines: $GW$+EDMFT ($N_{c}=4$). Pentagons:
determinant QMC (only a subset of $\mathbf{K}$
points is shown for better visibility).\label{fig:Sigma_weak_coupling}}
\end{figure}

\begin{figure}
\begin{centering}
\includegraphics[width=1\columnwidth]{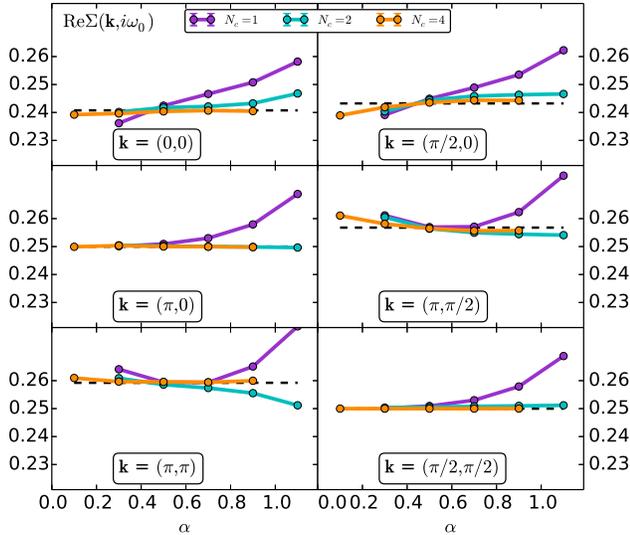} 
\par\end{centering}
\caption{Dependence of $\mathrm{Re}\Sigma(\mathbf{k},i\omega_{0})$ on $\alpha$ for different momenta 
 (Point A: $U/D=0.5$, $\delta=0\%$, $\beta D=16$, $t'=0$). Black dashed lines: QMC.\label{fig:alpha_dependence}}
\end{figure}

\begin{figure}
\begin{centering}
\includegraphics[width=1\columnwidth]{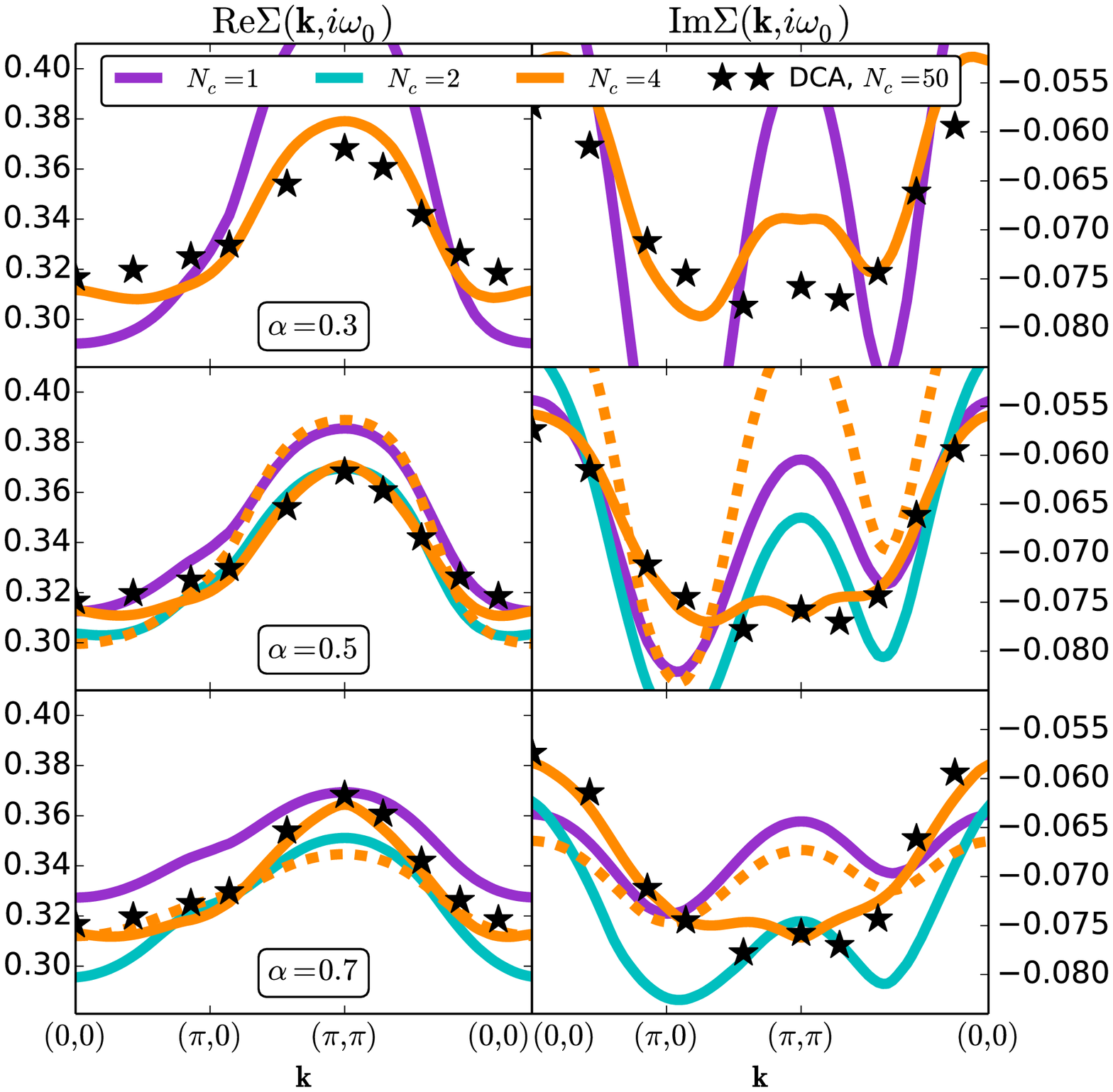} 
\par\end{centering}
\caption{$\Sigma(\mathbf{k},i\omega_{0})$ at point B ($U/D=1$, $\delta=20\%$, $\beta D=16$, $t'=0$). Same conventions
as Fig. \ref{fig:Sigma_weak_coupling}. Dashed lines: $GW$+EDMFT. Stars: DCA from Ref.~\protect\onlinecite{Leblanc2015}. \label{fig:Sigma_interm_coupling}}
\end{figure}

\begin{figure}
\begin{centering}
\includegraphics[width=1\columnwidth]{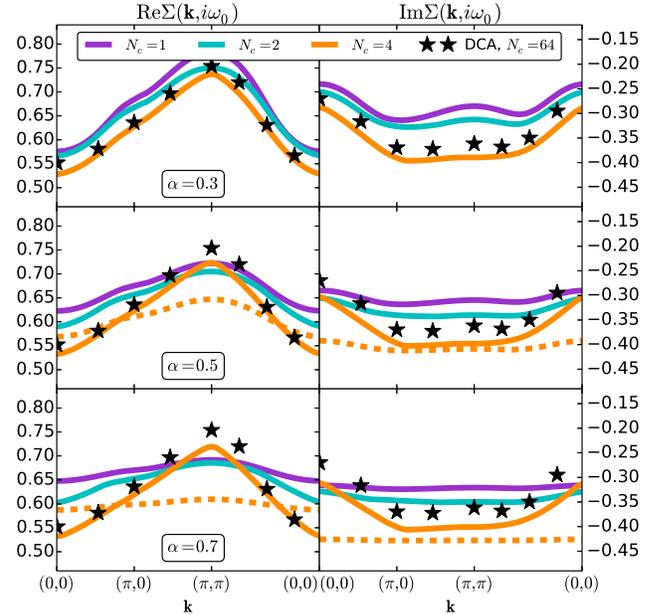} 
\par\end{centering}
\caption{$\Sigma(\mathbf{k},i\omega_{0})$ at point C ($U/D=1.4$, $\delta=4\%$, $\beta D=8$, $t'/t=-0.3$). Same
conventions as Fig. \ref{fig:Sigma_weak_coupling}. Dashed lines: $GW$+EDMFT. Stars: DCA.\label{fig:Sigma_strong_coupling}}
\end{figure}

We start with point A. In Fig. \ref{fig:Sigma_weak_coupling}, we
show the self-energy $\Sigma(\mathbf{k},i\omega_{0})$ for cluster
sizes of $N_{c}=1$ (single-site), $2$ (dimer) and 4 (plaquette)
and for three different values of $\alpha$. 
As expected, 
the dependence on $\alpha$ becomes weaker as $N_{c}$ increases.
At $N_{c}=4$, the self-energy is almost independent on
$\alpha$. 
The $\alpha$-dependence for $N_c=1,2,4$ is further illustrated 
in Fig.~\ref{fig:alpha_dependence}:  
the $N_c=4$ results show an extended plateau which
is narrower or nonexistent for $N_c=1,2$. 

The benchmarks, using numerically exact determinant QMC\cite{Blankenbecler1981} computed with $N_{c}=16\times16$ sites,
are also presented on both 
Fig.~\ref{fig:Sigma_weak_coupling} and Fig.~\ref{fig:alpha_dependence}.
We observe a very good agreement between $N_c=4$ and the benchmark data, both for the real and imaginary parts of the self-energy,
which validates the Fierz criterion in this regime.
We also observe that for $\alpha=0.5$, the results are in agreement with the converged values regardless of $N_c$.
This can be understood by noticing that $\alpha=0.5$
corresponds to the values of $U^\eta$ used in the random phase approximation (RPA),
which is correct to second order in $U$.

Moreover, we compare our results with the self-energy obtained by the
$GW$+EDMFT\cite{Biermann2003,Sun2002,Sun2004,Ayral2012,Ayral2013,Biermann2014,Ayral2017}
method for $N_{c}=4$. $GW$+EDMFT can be regarded as a simplification
of TRILEX where the vertex corrections are neglected in the nonlocal self-energy
contribution.
This explains why the $GW$+EDMFT results are,
independently of $\alpha$, quite close to the single-site
TRILEX results: the vertex frequency and momentum dependences are weak
in the low-$U$ limit.
Besides, they are different from the cluster
TRILEX results and from the exact solution, except for the RPA
value of $\alpha$ ($\alpha=0.5$) where both methods give results
close to the exact solution.

At point B (Fig.~\ref{fig:Sigma_interm_coupling}), the agreement
between the benchmarks and the real and imaginary parts
of the self-energy, for all values of $\alpha$ (with more important
deviations for $\alpha=0.3$), is very good for $N_{c}=4$. Contrary
to the weak-coupling limit, no value of $\alpha$ in the single-site
case matches the exact solution. This points to the importance of
nonlocal corrections to the three-leg vertex. This observation is
further corroborated by looking at the $GW$+EDMFT curve. There, the
agreement with the exact result is quite poor, while being similar
to the single-site result, like in the weak-coupling limit (for $\alpha=0.3$,
a spin instability precludes convergence of $GW$+EDMFT and cluster TRILEX for $N_c=2$). This discrepancy
shows that as interactions are increased, the vertex frequency and momentum
dependence play a more and more important role in the
nonlocal self-energy, as we will discuss below. These conclusions
are also valid for local observables (see Suppl. Mat. C.3).

At the strong-coupling point C (Fig. \ref{fig:Sigma_strong_coupling}),
similarly to the previous regimes, the $N_{c}=4$ self-energy is almost
independent of $\alpha$, and in good agreement
with the converged (DCA) solution (especially for its real part).
$GW$+EDMFT at $N_{c}=4$ is quite far from the exact result, as can
be expected from the previous discussion.

\begin{figure}
\begin{centering}
\includegraphics[width=0.9\columnwidth]{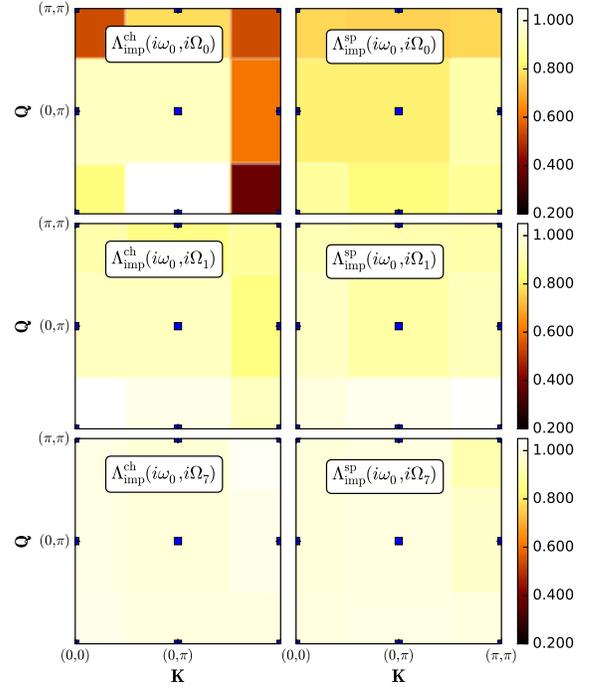} 
\par\end{centering}
\caption{Point B ($U/D=1$, $\delta=20\%$, $\beta D=16$, $t'=0$), $\alpha=0.5$.
Impurity vertex $\Lambda_{\mathrm{imp}}^{\eta}(\mathbf{K},\mathbf{Q};i\omega_{0},i\Omega)$
at $\mathbf{K},\mathbf{Q}\in[(0,0),(0,\pi),(\pi,\pi)]^{2}$ (the value
is color-coded in the square area surrounding each blue point) in
the charge (left) and spin (right) channels, for increasing
bosonic Matsubara frequency (from top to bottom). \label{fig:Impurity-vertex-interm-coupling}}
\end{figure}

Finally, we analyze the momentum and frequency dependence of the vertex, illustrated in Fig. \ref{fig:Impurity-vertex-interm-coupling}.
At low Matsubara frequencies, the vertex acquires a momentum dependence
(especially in the charge channel), while it is essentially local
at high frequencies. In other words, the largest deviations to locality
occur at small frequencies only (see also Suppl. Mat. C.4).
The nonlocal components are smaller or much smaller than the local
component, especially for large Matsubara frequencies. This gives
an \emph{a posteriori} explanation of the qualitatively good results
of the single-site TRILEX approximation. More importantly, the fact
that the momentum dependence is confined to low frequencies suggests
optimizations for the vertex parametrization and computation.

In conclusion, we have presented a first implementation of the cluster
extension of the TRILEX method. 
For a broad interaction and doping
range of the two-dimensional Hubbard model, we obtain, for an embedded
cluster with only four impurity sites, continuous self-energies in
close agreement with the exact result obtained with comparatively
expensive large-cluster lattice QMC and DCA calculations.

Cluster TRILEX is based on the computation and momentum coarse-graining of the three-leg vertex function: it thus comes at a cost lower than cluster methods based on four-leg vertices\cite{Hafermann2008,Slezak2009}, but it \emph{a priori} suffers from the Fierz ambiguity.
We have shown that this ambiguity can be turned into
a practical advantage in two ways: First and foremost, we have shown
that proximity to the exact solution coincides with stability with
respect to the Fierz parameter $\alpha$ \footnote{This also holds for other HS decouplings, Suppl. Mat. A}. With this necessary condition, one can assess,
at a given (possibly small) cluster size, the accuracy of the solution.
Second, in some regimes, there exists a value of $\alpha$
for which accurate results can be reached for smaller cluster sizes.
By allowing to extract more information from smaller embedded TRILEX clusters, the Fierz convergence criterion paves the way to a controlled exploration of low-temperature 
phases such as superconducting phases, where
cluster DMFT cannot be converged in practice. 

\begin{acknowledgments}
We acknowledge useful discussions with M. Ferrero and A. Georges.
We especially thank W. Wu for providing us determinant QMC numerical
data for the benchmark results of point A and DCA data for point C,
as well as J. LeBlanc for providing us the DCA data (from Ref. \onlinecite{Leblanc2015})
for point B. This work is supported by the FP7/ERC, under Grant Agreement
No. 278472-MottMetals. Part of this work was performed using HPC resources
from GENCI-TGCC (Grant No. 2016-t2016056112). Our implementation is
based on the TRIQS toolbox\cite{Parcollet2014}. 

\end{acknowledgments}

\clearpage
\appendix
\global\long\def\thefigure{S.\arabic{figure}}
 \setcounter{figure}{0} \global\long\def\appendixname{Supplemental Material}

\renewcommand\appendixname{Supplemental Material}

This Supplemental Material is organized as follows: in Section \ref{sec:Self-energy-Heisenberg},
we show results corresponding to another decoupling than the Ising
decoupling used in the main text, namely the Heisenberg decoupling.
In Section \ref{sec:Technical-details-of}, we give the technical
details relevant to the implementation of the cluster TRILEX method.
Finally, in Section \ref{sec:Supplementary-data}, we give supplementary
data to complement the figures and discussion of the main text.

\section{Self-energy in the Heisenberg decoupling: $\alpha$ and $N_{c}$
dependence and comparison to exact benchmarks\label{sec:Self-energy-Heisenberg}}

\begin{figure}
\begin{centering}
\includegraphics[width=1\columnwidth]{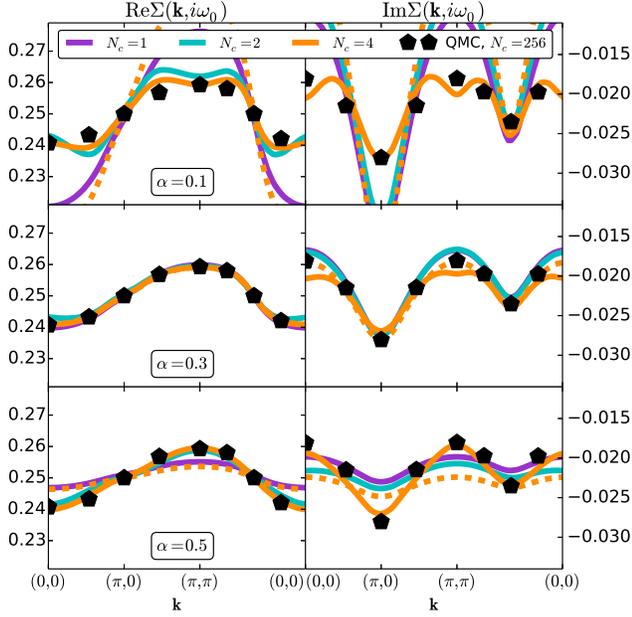} 
\par\end{centering}
\caption{Point A in the Heisenberg decoupling ($U/D=0.5$, $\delta=0\%$, $\beta D=16$).
$\mathrm{Re}\Sigma(\mathbf{k},i\omega_{0})$ (left column) and $\mathrm{Im}\Sigma(\mathbf{k},i\omega_{0})$
(right column) for $N_{c}=1,2,4$ for various values of the Fierz
parameter $\alpha$ (from top to bottom), along the path $(0,0)-(\pi,0)-(\pi,\pi)-(0,0)$.
Solid lines: TRILEX. Dashed lines: $GW$+EDMFT ($N_{c}=4$). Pentagons:
determinant QMC ($N_{c}=256$; only a small subset of $\mathbf{K}$
points is shown for a better visibility).\label{fig:Sigma_weak_coupling}}
\end{figure}

\begin{figure}
\begin{centering}
\includegraphics[width=1\columnwidth]{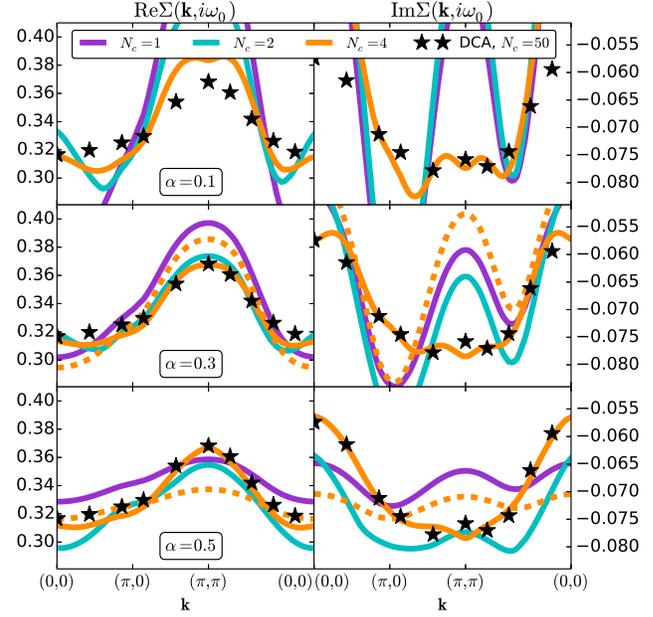} 
\par\end{centering}
\caption{Point B in the Heisenberg decoupling ($U/D=1$, $\delta=20\%$, $\beta D=16$).
Same conventions as Fig. \ref{fig:Sigma_weak_coupling}. Stars: DCA
from Ref.~\protect\protect\onlinecite{Leblanc2015}, $N_{c}=50$.
\label{fig:Sigma_interm_coupling}}
\end{figure}

\begin{figure}
\begin{centering}
\includegraphics[width=1\columnwidth]{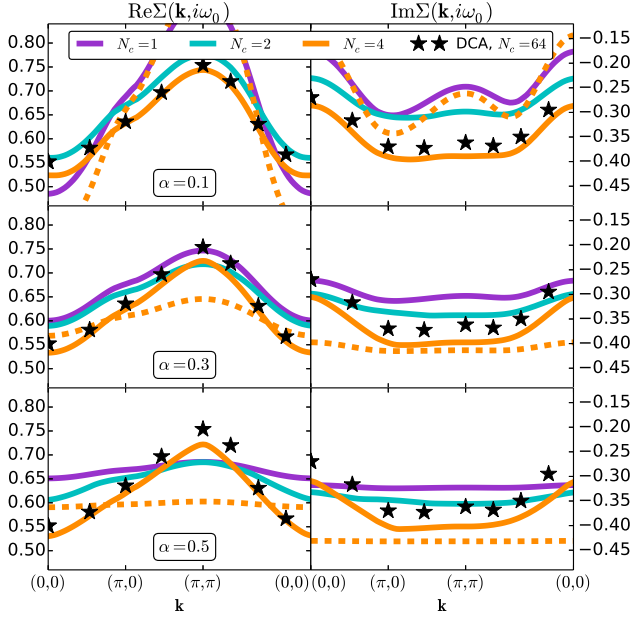} 
\par\end{centering}
\caption{Point C in the Heisenberg decoupling ($U/D=1.4$, $\delta=4\%$, $\beta D=8$).
Same conventions as Fig. \ref{fig:Sigma_weak_coupling}. Stars: DCA
from Ref.~\protect\protect\onlinecite{Wu2016}, $N_{c}=64$\label{fig:Sigma_strong_coupling}}
\end{figure}

In the main text, we have chosen to decouple the interaction with
charge and longitudinal spin bosons (a decoupling sometimes called
the ``Ising'' decoupling). One can alternatively use the ``Heisenberg''
decoupling, which consists in decomposing the interaction as follows
(up to a density term): 
\begin{equation}
Un_{i\uparrow}n_{i\downarrow}=\frac{1}{2}U^{\mathrm{ch}}n_{i}n_{i}+\frac{1}{2}U^{\mathrm{sp}}\left(s_{i}^{x}s_{i}^{x}+s_{i}^{y}s_{i}^{y}+s_{i}^{z}s_{i}^{z}\right)\label{eq:U_decoupling}
\end{equation}

where $s_{i}^{I}\equiv\sum_{\sigma\sigma'}c_{i\sigma}^{\dagger}\sigma_{\sigma\sigma'}^{I}c_{i\sigma'}$
(with $\sigma^{I}$ the Pauli matrices). This equality holds whenever
$U^{\mathrm{ch}}-3U^{\mathrm{sp}}=U$, or in other words

\begin{align}
U^{\mathrm{ch}} & =(3\alpha-1)U,\;\;U^{\mathrm{sp}}=(\alpha-2/3)U\label{eq:U_Fierz}
\end{align}

This leads, after a Hubbard-Stratonovich transformation, to four bosonic
modes, one in the charge channel and three in the spin channel (we
refer the reader to \cite{Ayral2015c} for more details and for the
modified equations for the self-energy and impurity action).

In Figs (\ref{fig:Sigma_weak_coupling}-\ref{fig:Sigma_interm_coupling}-\ref{fig:Sigma_strong_coupling}),
we show the self-energies obtained for the three characteristic points
studied in the main text (A, B and C) for different values of the
Fierz parameter $\alpha$ and cluster size $N_{c}$.

The observations with respect to $\alpha$ dependence are very
similar to those made in the main text. This further underlines the
main conclusion of the paper: even in this quite different decoupling,
the results are similar to those obtained within the Ising decoupling
of the main text.

\section{Technical details of cluster TRILEX\label{sec:Technical-details-of}}

\subsection{Fourier conventions and patching details\label{sec:Fourier-conventions}}

\begin{figure}
\begin{centering}
\includegraphics[width=0.8\columnwidth]{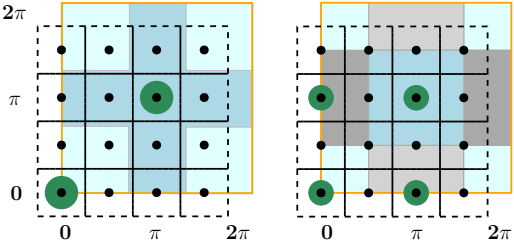} 
\par\end{centering}
\caption{Example of discretization of the Brillouin zone with $n_{\mathrm{latt}}=n_{k}\times n_{k}$
$\mathbf{k}$ points (here $n_{\mathbf{k}}=4$) for $N_{c}=2$ (left
panel) and $N_{c}=4$ (right panel)\label{fig:Example-of-discretization} }
\end{figure}

\subsubsection{Spatial Fourier transforms}

$\mathbf{k}$ is a Brillouin zone momentum (black dots in Fig. \ref{fig:Example-of-discretization}).

\paragraph{Direct transforms}

We define:

\begin{align}
f_{\mathbf{k}} & \equiv\sum_{i=1}^{n_{\mathrm{latt}}}e^{-i\mathbf{k}\cdot\mathbf{r}_{i}}f_{\mathbf{r}_{i}}\label{eq:FT_def_onevar}
\end{align}

\paragraph{Reciprocal transforms}

We define:

\begin{align}
f_{\mathbf{r}} & =\frac{1}{n_{\mathrm{latt}}}\sum_{i=1}^{n_{\mathrm{latt}}}e^{i\mathbf{k}_{i}\cdot\mathbf{r}}f_{\mathbf{k}}\label{eq:recip_FT_space}
\end{align}

\subsubsection{Cluster Fourier transforms}

$\mathbf{K}$ and $\mathbf{Q}$ are cluster momenta (green disks in
Fig. \ref{fig:Example-of-discretization})

\paragraph{Direct transforms}

We define:

\begin{align}
f_{\mathbf{K}} & \equiv\frac{1}{N_{c}}\sum_{ij}e^{-i\mathbf{K}\cdot(\mathbf{R}_{i}-\mathbf{R}_{j})}f_{ij}\label{eq:FT_def_onevar-cluster}\\
g_{\mathbf{K},\mathbf{Q}} & \equiv\frac{1}{N_{c}}\sum_{ijk}e^{-i\mathbf{K}\cdot(\mathbf{R}_{i}-\mathbf{R}_{j})-i\mathbf{Q}\cdot(\mathbf{R}_{k}-\mathbf{R}_{j})}g_{ijk}\label{eq:FT_def_twovar-cluster}
\end{align}
with $i,j,k=1\dots N_{c}$.

\paragraph{Reciprocal transforms}

We define:

\begin{align}
f_{ij} & =\sum_{\mathbf{K}}e^{i\mathbf{K}\cdot(\mathbf{R}_{i}-\mathbf{R}_{j})}f_{\mathbf{K}}\label{eq:recip_FT_cluster}\\
g_{ijk} & =\sum_{\mathbf{K}\mathbf{Q}}e^{i\mathbf{K}(\mathbf{R}_{i}-\mathbf{R}_{j})+i\mathbf{Q}(\mathbf{R}_{k}-\mathbf{R}_{j})}g_{\mathbf{K},\mathbf{Q}}\label{eq:recip_FT_cluster_twovar}
\end{align}

where $\sum_{\mathbf{K}}f_{\mathbf{K}}$ is shorthand for $\frac{1}{N_{c}}\sum_{i=1}^{N_{c}}f_{\mathbf{K}_{i}}$.

\subsubsection{Temporal Fourier transforms}

$i\omega$ (resp. $i\Omega)$ denotes fermionic (resp. bosonic) Matsubara
frequencies, and are shorthand for $i\omega_{n}=\frac{2n+1}{\beta}\pi$
(resp. $i\Omega_{m}=\frac{2m}{\beta}\pi$). $\beta$ is the inverse
temperature.

\paragraph{Direct transforms}

We define:

\begin{align}
f_{i\omega} & \equiv\int_{0}^{\beta}\mathrm{d}\tau e^{i\omega\tau}f_{\tau}\label{eq:FT_def_onevar-time}\\
g_{i\omega,i\Omega} & \equiv\iint_{0}^{\beta}\mathrm{d\tau}\mathrm{d}\tau'e^{i\omega\tau+i\Omega\tau'}g_{\tau,\tau'}\label{eq:FT_def_twovar-time}
\end{align}

\paragraph{Reciprocal transforms}

We define:

\begin{align}
f_{\tau} & =\sum_{i\omega}e^{-i\omega\tau}f_{i\omega}\label{eq:recip_FT_onevar_time}\\
g_{\tau,\tau'} & =\sum_{i\omega}\sum_{i\Omega}e^{-i\omega\tau-i\Omega\tau'}g_{i\omega,i\Omega}\label{eq:recip_FT_twovar_time}
\end{align}

Here, $\sum_{i\omega}f(i\omega)$ is shorthand for $\frac{1}{\beta}\sum_{n=-n_{\mathrm{max}}}^{n_{\mathrm{max}}-1}f(i\omega_{n})$
(and $\sum_{i\Omega}f(i\Omega)$ for $\frac{1}{\beta}\sum_{m=-m_{\mathrm{max}}}^{m_{\mathrm{max}}}f(i\Omega_{m})$).

\subsubsection{Patching and discretization\label{subsec:Patching}}

In DCA, the $\mathbf{k}$ integrals can be replaced with integrals
on the density of states, e.g. 
\begin{align*}
G_{\mathbf{K}}(i\omega) & =\sum_{\mathbf{k}\in\mathcal{P}_{\mathbf{K}}}\frac{1}{i\omega+\mu-\varepsilon_{\mathbf{k}}-\Sigma_{\mathrm{imp}}(\mathbf{K},i\omega)}\\
 & =\int_{-\infty}^{\infty}\mathrm{d}\varepsilon\frac{D_{\mathbf{K}}(\varepsilon)}{i\omega+\mu-\varepsilon-\Sigma_{\mathrm{imp}}(\mathbf{K},i\omega)}
\end{align*}

where $D_{\mathbf{K}}(\varepsilon)\equiv\sum_{\mathbf{k}\in\mathcal{P}_{\mathbf{K}}}\delta(\varepsilon-\varepsilon_{\mathbf{k}})$
is the noninteracting density of states of patch $\mathbf{K}$. This
density of states can be precomputed once and for all for a given
dispersion and patches with a very large number of $\mathbf{k}$ points
to obtain a very good accuracy.

By contrast, in cluster TRILEX, the self-energy is a function of $\mathbf{k}$
instead of $\mathbf{K}$, forbidding this substitution and keeping
the number of $\mathbf{k}$ points finite (this number is primarily
limited by memory and computation time requirements, but it can be
large due to the low cost of the computation of $\Sigma(\mathbf{k},i\omega)$:
we typically discretize the Brillouin zone in $n_{\mathbf{k}}\times n_{\mathbf{k}}$
points, with $n_{\mathbf{k}}=32$).

This requires extra care when defining the theta functions $\theta_{\mathbf{K}}(\mathbf{k})$
defined in a loose way in the main text. $\theta_{\mathbf{K}}(\mathbf{k})$
is precisely defined as the overlap of the area surrounding a given
$\mathbf{k}$ point with the patch $\mathcal{P}_{\mathbf{K}}$, divided
by the total area surrounding the $\mathbf{k}$ point. This area is
illustrated in Fig. \ref{fig:Example-of-discretization} for the case
$n_{\mathbf{k}}=4$. For instance, the $\mathbf{k}$ point of coordinates
$(1,1)$ has $\theta_{\mathbf{K}=(0,0)}(\mathbf{k})=1/4$, while that
of coordinates $(1,2)$ has $\theta_{\mathbf{K}=(0,\pi)}(\mathbf{k})=1/2$.

Correspondingly, $\sum_{\mathbf{k}\in\mathcal{P}_{\mathbf{K}}}$ is
precisely defined as 
\begin{equation}
f_{\mathbf{K}}=\sum_{\mathbf{k}\in\mathcal{P}_{\mathbf{K}}}f_{\mathbf{k}}=\frac{\sum_{i=1}^{n_{\mathbf{k}}\times n_{\mathbf{k}}}f(\mathbf{k}_{i})\theta_{\mathbf{K}}(\mathbf{k}_{i})}{\sum_{i=1}^{n_{\mathbf{k}}\times n_{\mathbf{k}}}\theta_{\mathbf{K}}(\mathbf{k}_{i})}\label{eq:patching_definition}
\end{equation}

\subsection{Cluster TRILEX Loop\label{sec:Trilex-Loop}}

\begin{figure}
\begin{centering}
\includegraphics[width=0.85\columnwidth]{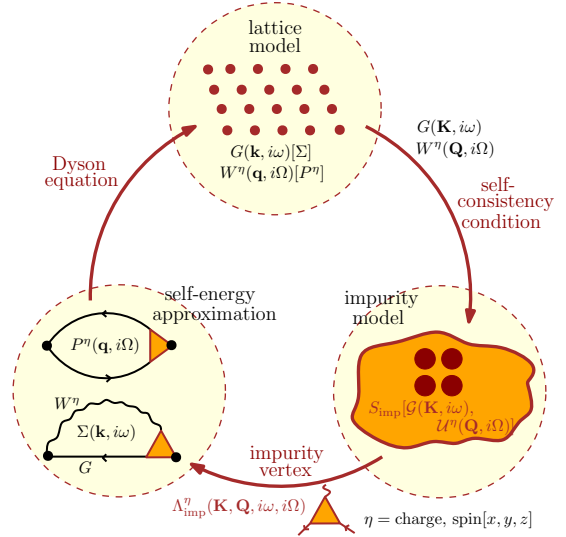} 
\par\end{centering}
\caption{The cluster TRILEX loop\label{fig:The-cluster-TRILEX-loop}}
\end{figure}

As in Refs~\onlinecite{Ayral2015, Ayral2015c, Vucicevic2017}, we
solve the cluster TRILEX equations by forward recursion, with the
following steps (illustrated in Fig. \ref{fig:The-cluster-TRILEX-loop}): 
\begin{enumerate}
\item Start with a guess $\Sigma(\mathbf{k},i\omega)$, $P^{\eta}(\mathbf{q},i\Omega)$ 
\item Compute $G(\mathbf{k},i\omega)$ and $W^{\eta}(\mathbf{q},i\Omega)$
(Eqs (4)) and then $G(\mathbf{K},i\omega)$ and $W^{\eta}(\mathbf{Q},i\Omega)$
(Eqs. (10)) 
\item Compute $\mathcal{G}(\mathbf{K},i\omega)$ and $\mathcal{U}^{\eta}(\mathbf{Q},i\Omega)$
by substituting Eqs (9) into the impurity Dyson equations, i.e\begin{subequations}
\begin{align}
\mathcal{G}(\mathbf{K},i\omega) & =\left[G_{\mathbf{K}}^{-1}(i\omega)+\Sigma_{\mathrm{imp}}(\mathbf{K},i\omega)\right]^{-1}\label{eq:G_weiss}\\
\mathcal{U}^{\eta}(\mathbf{Q},i\Omega) & =\left[\left[W_{\mathbf{Q}}^{\eta}\right]^{-1}(i\Omega)+P_{\mathrm{imp}}^{\eta}(\mathbf{Q},i\Omega)\right]^{-1}\label{eq:U_weiss}
\end{align}
\end{subequations} 
\item Solve the impurity model, Eq. (6), for its exact vertex $\Lambda_{\mathrm{imp}}^{\eta}(\mathbf{K},\mathbf{Q};i\omega,i\Omega)$
(see Section \ref{sec:Impurity} for more details). 
\item Compute $\Sigma(\mathbf{k},i\omega)$ and $P^{\eta}(\mathbf{q},i\Omega)$
(Eqs (5)) 
\item Go back to step 2 until convergence of $\Sigma$ and $P^{\eta}$. 
\end{enumerate}
As in Refs~\onlinecite{Ayral2015, Vucicevic2017}, and as justified
in Ref.~\onlinecite{Ayral2015c} for the single-site impurity case,
in the equations presented in the main text and in the loop presented
above, we have implicitly approximated the impurity's electron-boson
vertex with the bare electron-boson vertex or, in other words, we
have assumed the $\zeta$ function, introduced in Ref.~\onlinecite{Ayral2015c},
to be negligible.

\subsection{Solution of the Impurity Model\label{sec:Impurity}}

\subsubsection{Impurity solver\label{subsec:Impurity-solver}}

The impurity model, defined by Eq. (6), is solved using a continuous-time
quantum Monte-Carlo algorithm\cite{Rubtsov2005}. For $N_c=1$, we refer the reader to Ref.~\onlinecite{Ayral2015c} for details. For $N_c>1$, contrary to the
single-site case, the densities $n_{i}^{I}$ are no longer good quantum
numbers due to the intra-cluster hopping terms. This precludes the
use of the hybridization expansion algorithms, which can be used with
retarded interactions only if the operators involved in the retarded
interactions are good quantum numbers, and in which only correlators
between operators which are good quantum numbers can be easily measured.
We therefore use an interaction-expansion (CT-INT) algorithm, described
\emph{e.g.} in Ref.~\onlinecite{Gull2011}. Here, for the measurement
of the three-point function $\tilde{\chi}_{\mathrm{imp}}^{3,\sigma\sigma'}(i,j,k;\tau,\tau')$
(defined in Eq.~\eqref{eq:chi3_imp_def} below), we use a straightforward
operator-insertion method.

\begin{figure}
\begin{centering}
\includegraphics[width=1\columnwidth]{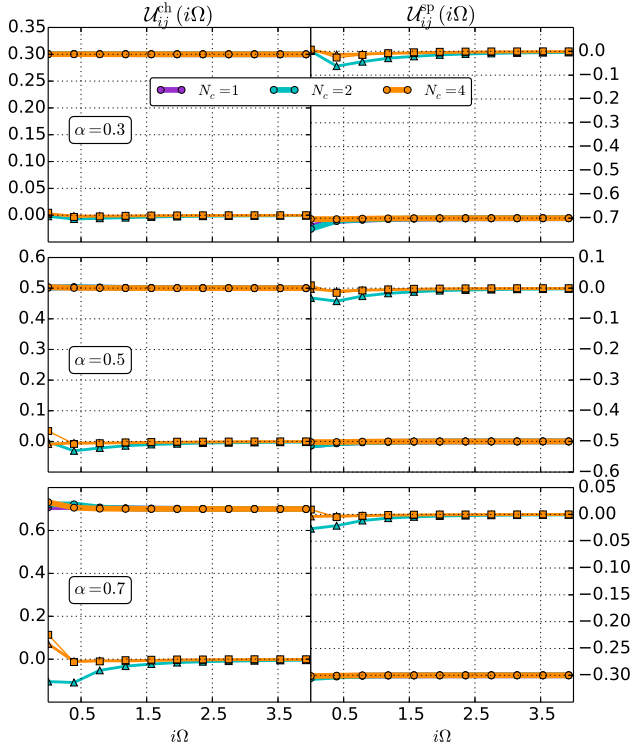} 
\par\end{centering}
\caption{Retarded interaction $\mathcal{U}_{ij}^{\eta}(i\Omega)$ in the charge
(left column) and spin (right column) channels, for $\alpha=0.3$
(top row), $0.5$ (middle row), $0.7$ (bottom row), at point B ($U/D=1$,
$\delta=20\%$, $\beta D=16$, $t'=0$, Ising decoupling). Dots: local
component ($i,j=0,0$). Triangles: nearest-neighbor component ($i,j=0,1$,
for $N_{c}=2$ and $N_{c}=4$ only). Squares: next-nearest-neighbor
component ($i,j=0,3$, for $N_{c}=4$ only). \label{fig:Retarded-interaction}}
\end{figure}

We observe that in all the parameter regimes studied in the main text
(points A, B and C), the interactions $\mathcal{U}_{ij}^{I}(\tau)$
are static and local to a very good approximation: 
\begin{equation}
\mathcal{U}_{ij}^{I}(\tau)\approx U^{I}\delta_{ij}\delta_{\tau}\label{eq:static_interactions}
\end{equation}
This is illustrated in Fig. \ref{fig:Retarded-interaction} for point
B. Thus, in practice, we do not have to use the retarded interactions.
This simplifies the numerical computation since the dependence of
the Monte-Carlo sign problem on CT-INT's density-shifting parameter
$\alpha_{\sigma}(s)$ (see e.g. Eq. (145) of Ref.~\onlinecite{Gull2011})
is less simple than in the case of static interactions.

\subsubsection{Computation of $G_{\mathrm{imp}}(\mathbf{K},i\omega)$ and $W_{\mathrm{imp}}(\mathbf{Q},i\Omega)$\label{subsec:G_imp_W_imp}}

$G_{\mathrm{imp}}(\mathbf{K},i\omega)$ and $W_{\mathrm{imp}}(\mathbf{Q},i\Omega)$
are obtained by computing the spatial and temporal Fourier transforms
(defined in Section \ref{sec:Fourier-conventions}) $G_{\mathrm{imp}}(\mathbf{K},i\omega)$
and $\chi_{\mathrm{imp}}^{\sigma\sigma'}(\mathbf{Q},i\Omega)$ of
the impurity's Green's function and density-density response functions:
\begin{subequations} 
\begin{align}
G_{\mathrm{imp}}(i,j;\tau) & \equiv-\langle Tc_{i}(\tau)c_{j}^{\dagger}(0)\rangle_{\mathrm{imp}}\label{eq:G_imp_def}\\
\chi_{\mathrm{imp}}^{\sigma\sigma'}(i,j;\tau) & \equiv\langle Tn_{i\sigma}(\tau)n_{j\sigma'}(0)\rangle_{\mathrm{imp}}\label{eq:chi_imp_def}
\end{align}

\end{subequations}and by using the identity 
\begin{align}
 & W_{\mathrm{imp}}^{\eta}(\mathbf{Q},i\Omega)=\label{eq:W_chi_rel}\\
 & \mathcal{U}^{\eta}(\mathbf{Q},i\Omega)-\mathcal{U}^{\eta}(\mathbf{Q},i\Omega)\chi_{\mathrm{imp}}^{\eta,\mathrm{conn}}(\mathbf{Q},i\Omega)\mathcal{U}^{\eta}(\mathbf{Q},i\Omega)\nonumber 
\end{align}
where the passage from spin ($\sigma,\sigma'$) to channel ($\eta$)
indices is done using the expressions: \begin{subequations} 
\begin{align}
\chi_{\mathrm{imp}}^{\eta=\mathrm{ch}} & \equiv2(\chi_{\mathrm{imp}}^{\uparrow\uparrow}+\chi_{\mathrm{imp}}^{\uparrow\downarrow})\label{eq:chi_ch}\\
\chi_{\mathrm{imp}}^{\eta=\mathrm{sp}} & \equiv2(\chi_{\mathrm{imp}}^{\uparrow\uparrow}-\chi_{\mathrm{imp}}^{\uparrow\downarrow})\label{eq:chi_sp}
\end{align}
\end{subequations} and the connected component is:

\begin{equation}
\chi_{\mathrm{imp}}^{\eta,\mathrm{conn}}(i,j;i\Omega)\equiv\chi_{\mathrm{imp}}^{\eta,\mathrm{conn}}(i,j;i\Omega)-\langle n_{i}^{\eta}\rangle\langle n_{j}^{\eta}\rangle\beta\delta_{i\Omega}
\end{equation}

\subsubsection{Computation of the cluster vertex $\Lambda_{\mathrm{imp}}^{\eta}(\mathbf{K},\mathbf{Q};i\omega,i\Omega)$\label{subsec:Lambda_imp}}

The computation of $\Lambda_{\mathrm{imp}}^{\eta}(\mathbf{K},\mathbf{Q};i\omega,i\Omega)$
is done by measuring the three-point function

\begin{equation}
\tilde{\chi}_{\mathrm{imp}}^{3,\sigma,\sigma'}(i,j,k;\tau,\tau')\equiv\langle Tc_{i\sigma}(\tau)c_{j\sigma}^{\dagger}(0)n_{k\sigma'}(\tau')\rangle_{\mathrm{imp}}\label{eq:chi3_imp_def}
\end{equation}

The vertex, written in cluster coordinates $\mathbf{R}_{i},\mathbf{R}_{j},\mathbf{R}_{k}$,
is then computed as:

\begin{align}
& \Lambda^{\eta}(i,j,k;i\omega,i\Omega) \equiv\sum_{pqr}G_{\mathrm{imp}}^{-1}(p,j;i\omega+i\Omega)G_{\mathrm{imp}}^{-1}(i,q;i\omega)\nonumber \\
 & \;\;\times\left[1-\mathcal{U}^{\eta}\chi_{\mathrm{imp}}^{\eta}\right]^{-1}(k,r; i\Omega)\tilde{\chi}_{\mathrm{imp}}^{3,\eta,\mathrm{conn}}(q,p,r;i\omega,i\Omega)\label{eq:vertex_def}
\end{align}

with the expression in the charge and spin channel:\begin{subequations}
\begin{align}
 & \tilde{\chi}_{\mathrm{imp}}^{3,\eta=\mathrm{ch}}\equiv\tilde{\chi}_{\mathrm{imp}}^{3,\uparrow\uparrow}+\tilde{\chi}_{\mathrm{imp}}^{3,\uparrow\downarrow}\label{eq:chi3_ch}\\
 & \tilde{\chi}_{\mathrm{imp}}^{3,\eta=\mathrm{sp}}\equiv\tilde{\chi}_{\mathrm{imp}}^{3,\uparrow\uparrow}-\tilde{\chi}_{\mathrm{imp}}^{3,\uparrow\downarrow}\label{eq:chi3_sp}
\end{align}
\end{subequations}and the connected component defined as:

\begin{align}
 & \tilde{\chi}_{\mathrm{imp}}^{3,\eta,\mathrm{conn}}(i,j,k;i\omega,i\Omega)\equiv\label{eq:chi3_conn}\\
 & \;\;\tilde{\chi}_{\mathrm{imp}}^{3,\eta}(i,j,k;i\omega,i\Omega)+G_{\mathrm{imp}}(i,j;i\omega)n_{k}^{\eta}\beta\delta_{i\Omega}\nonumber 
\end{align}

$\Lambda^{\eta}(i,j,k;i\omega,i\Omega)$ is then Fourier-transformed
to $\Lambda_{\mathrm{imp}}^{\eta}(\mathbf{K},\mathbf{Q};i\omega,i\Omega)$
(see Section \ref{sec:Fourier-conventions}, Eq. (\ref{eq:FT_def_twovar-cluster})).

In practice, instead of directly performing a temporal Fourier transform
to compute $\tilde{\chi}_{\mathrm{imp}}^{3,\sigma\sigma'}(i,j,k;i\omega,i\Omega)$
from $\tilde{\chi}_{\mathrm{imp}}^{3,\sigma\sigma'}(i,j,k;\tau,\tau')$,
we first compute the connected component $\tilde{\chi}_{\mathrm{imp}}^{3,\eta,\mathrm{conn}}(i,j,k;\tau,\tau')$
{[}defined in Eq. (\ref{eq:chi3_conn}){]}, which is smooth and without
discontinuities, perform a cubic spline interpolation of it, and then
Fourier transform it to Matsubara frequencies. This allows us to use
a small number (typically $n_{\tau}=n_{\tau'}=100$) of $\tau,\tau'$
points in the measurement.

\subsection{Self-energy decomposition\label{sec:Self-energy}}

In this section, we show that the coarse-grainings introduced for
the vertex allow for a numerically convenient decomposition of $\Sigma$
and $P$.

Following a procedure very similar to that described in section II.D.3
of Ref.~\onlinecite{Ayral2015c}, we decompose Eqs (5) as follows:\begin{subequations}

\begin{eqnarray}
 &  & \Sigma(\mathbf{k},i\omega)=\Sigma_{\mathrm{imp}}(i,j=0,0;i\omega)\label{eq:Sigma_k_DCA_decomp}\\
 &  & \;\;\;-\sum_{\eta}m_{\eta}\sum_{\mathbf{K},\mathbf{Q}}\sum_{\mathbf{q},i\Omega}\tilde{G}_{\mathbf{k}+\mathbf{q},i\omega+i\Omega}^{\mathbf{K}+\mathbf{Q}}\tilde{W}_{\mathbf{q},i\Omega}^{\eta,\mathbf{Q}}\Lambda_{\mathrm{imp}}^{\eta}(\mathbf{K},\mathbf{Q};i\omega,i\Omega)\nonumber \\
 &  & P^{\eta}(\mathbf{q},i\Omega)=P_{\mathrm{imp}}^{\eta}(i,j=0,0;i\Omega)\label{eq:P_Q_DCA_decomp}\\
 &  & \;\;\;+2\sum_{\mathbf{K},\mathbf{Q}}\sum_{\mathbf{k},i\omega}\tilde{G}_{\mathbf{k}+\mathbf{q},i\omega+i\Omega}^{\mathbf{K}+\mathbf{Q}}\tilde{G}_{\mathbf{k},i\omega}^{\mathbf{K}}\Lambda_{\mathrm{imp}}^{\eta}(\mathbf{K},\mathbf{Q};i\omega,i\Omega)\nonumber 
\end{eqnarray}
\end{subequations}where we have defined the nonlocal components:
\begin{equation}
\tilde{X}(\mathbf{k},i\omega)\equiv X(\mathbf{k},i\omega)-\sum_{\mathbf{k}}X(\mathbf{k},i\omega)\label{eq:Xtilde_def}
\end{equation}
with $X=G$ or $W$.

Indeed, decomposing Eq. (5a) using Eq. (\ref{eq:Xtilde_def}), and
expanding, one obtains four terms, two of which vanish. The two remaining
terms are given in Eq. (\ref{eq:Sigma_k_DCA_decomp}). The first term
is given by $\Sigma_{\mathrm{imp}}(00,i\omega)$:

\begin{align}
 & -\sum_{\eta}m_{\eta}\sum_{\mathbf{K},\mathbf{Q}}\sum_{i\Omega}\left\{ \sum_{\mathbf{k}'}G_{i\omega+i\Omega}(\mathbf{k}')\theta_{\mathbf{K}+\mathbf{Q}}(\mathbf{k}')\right\} \nonumber \\
 & \;\;\times\left\{ \sum_{\mathbf{q}'}W_{i\Omega}(\mathbf{q}')\theta_{\mathbf{Q}}(\mathbf{q}')\right\} \Lambda_{\mathrm{imp}}^{\eta}(\mathbf{K},\mathbf{Q};i\omega,i\Omega)\\
 & =-\sum_{\eta}m_{\eta}\sum_{i\Omega}\sum_{\mathbf{q}'}\sum_{\mathbf{k}'}\left\{ G_{i\omega+i\Omega}(\mathbf{k}'+\mathbf{q}')\right\} \left\{ W_{i\Omega}(\mathbf{q}')\right\} \nonumber \\
 & \;\;\times\sum_{\mathbf{K}\mathbf{Q}}\theta_{\mathbf{K}+\mathbf{Q}}(\mathbf{k}'+\mathbf{q}')\theta_{\mathbf{Q}}(\mathbf{q}')\Lambda_{\mathrm{imp}}^{\eta}(\mathbf{K},\mathbf{Q};i\omega,i\Omega)\\
 & =-\sum_{\mathbf{k}'}\sum_{\eta}m_{\eta}\sum_{\mathbf{q}'}\sum_{i\Omega}G_{i\omega+i\Omega}(\mathbf{k}'+\mathbf{q}')W_{i\Omega}(\mathbf{q}')\Lambda_{\mathbf{k}',\mathbf{q}'}^{\eta}(i\omega,i\Omega)\nonumber \\
 & =\sum_{\mathbf{k}'}\Sigma(\mathbf{k}',i\omega)\nonumber \\
 & =\Sigma(\mathbf{R}=0,i\omega)\nonumber \\
 & =\Sigma_{\mathrm{imp}}(0,0;i\omega)\label{eq:Sigma_loc_Sigma_imp}
\end{align}

A similar result holds for $P$.

In the second terms of Eqs (\ref{eq:Sigma_k_DCA_decomp}-\ref{eq:P_Q_DCA_decomp}),
the summands decay fast for large Matsubara frequencies thanks to
the fast decay of the nonlocal component $\tilde{G}(\mathbf{k},i\omega)$
and $\tilde{W}(\mathbf{q},i\Omega)$.

As in Ref.~\onlinecite{Ayral2015c}, we furthermore split $\Lambda$
into a ``regular part'' $\Lambda^{\eta,\mathrm{reg}}$ which vanishes
at large frequencies

\begin{equation}
\Lambda^{\eta,\mathrm{reg}}(i,j,k;i\omega,i\Omega)=\Lambda^{\eta}(i,j,k;i\omega,i\Omega)-l^\eta(i,j,k;i\Omega)\label{eq:Lambda_reg_def}
\end{equation}
and a remainder $l^\eta(i\Omega)$ corresponding to the high-frequency
asymptotics of the three-point function: 
\begin{equation}
l^\eta(i,j,k;i\Omega)\equiv\sum_{p}\left[1-\mathcal{U}^{\eta}\chi^{\eta}\right]^{-1}(k,p;i\Omega)\delta_{ij}\label{eq:residuum}
\end{equation}

The term containing $\Lambda^{\eta,\mathrm{reg}}(i,j,k;i\omega,i\Omega)$
has a quickly decaying summand thanks to $\tilde{G}$, $\tilde{W}$
\emph{and} $\Lambda^{\mathrm{reg}}$. We compute it in Matsubara frequencies
 and real space after a fast Fourier transform of $\tilde{G}$
and $\tilde{W}$ (see Eq (\ref{eq:recip_FT_space})). This is the
bottleneck of the computation of the self-energy as it scales as $O(N_{\omega}^{2}N_{k}\log N_{k}N_{c}^{2})$
(where $N_{\omega}$ is the number of Matsubara frequencies used and
$N_{k}$ the number of $\mathbf{k}$ points in the disctretized first
Brillouin zone). The term containing $l^\eta(i,j,k;i\Omega)$ can be computed
entirely in imaginary time and real space, with a computational complexity
of $O(N_{\omega}\log N_{\omega}N_{k}\log N_{k}N_{c}^{2})$.

\section{Supplementary data\label{sec:Supplementary-data}}

\subsection{Additional data for the Fierz criterion: $\alpha$-dependence of
$\mathrm{Im}\Sigma$}

\begin{figure}
\begin{centering}
\includegraphics[width=1\columnwidth]{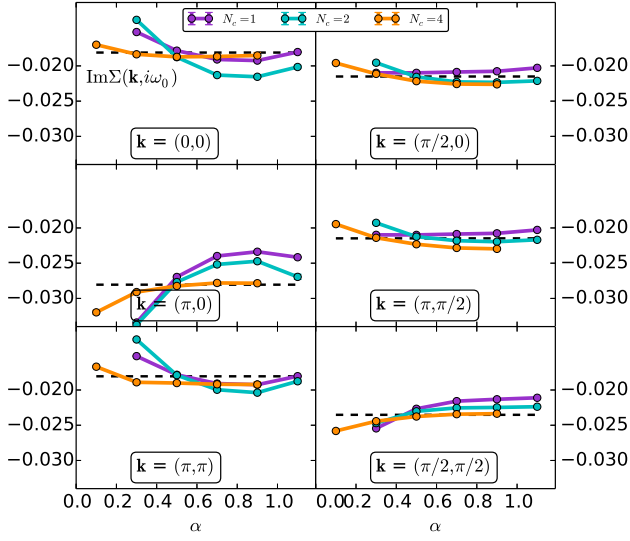} 
\par\end{centering}
\caption{Dependence of $\mathrm{Im}\Sigma(\mathbf{k},i\omega_{0})$ on the
Fierz parameter $\alpha$ for different $\mathbf{k}$ points at point
A ($U/D=0.5$, $\delta=0\%$, $\beta D=16$, $t'=0$, Ising decoupling).
\label{fig:alpha_dependence_Im}}
\end{figure}

In Figure \ref{fig:alpha_dependence_Im}, we complement the data of
Fig. 3 of the main text by giving the data for the imaginary part.
Similarly to the real part, the imaginary part shows plateaus for
given ranges of $\alpha$ which are more pronounced for $N_{c}=4$,
which is the cluster size for which the self-energy is the closest
to the exact benchmark result.

\subsection{Continuity of the self-energy\label{subsec:Continuity-of-the}}

\begin{figure}
\begin{centering}
\includegraphics[width=0.85\columnwidth]{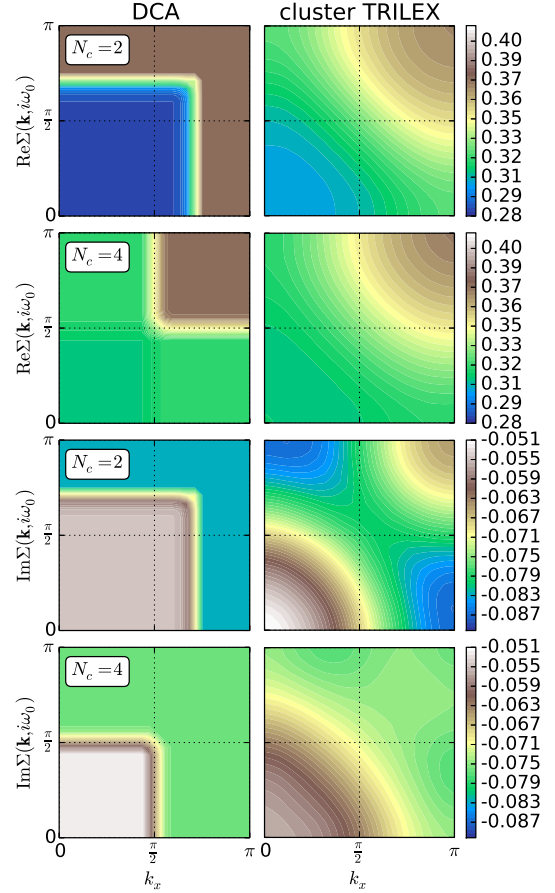} 
\par\end{centering}
\caption{$\Sigma(\mathbf{k},i\omega_{0})$ in the upper quadrant of the first
Brillouin zone, at point B ($U/D=1$, $\delta=20\%$, $\beta D=16$,
$t'=0$, $\alpha=0.5$, Ising decoupling). \emph{Left column}: DCA,
\emph{right column}: cluster TRILEX. \emph{First two rows}: real part,
\emph{last two rows}: imaginary part. \emph{Odd rows}: $N_{c}=2$,
\emph{even rows}: $N_{c}=4$.\label{fig:Sigma_colorplot}}
\end{figure}

In Fig. \ref{fig:Sigma_colorplot}, we show the lowest Matsubara component
of the self-energy obtained in the dynamical cluster approximation
(DCA) and the one obtained within cluster TRILEX, using Eq. (\ref{eq:Sigma_k_DCA_decomp}).
While the DCA self-energy is piecewise constant in the Brillouin zone
(with discontinuities at the patch edges), the cluster TRILEX self-energy
is continuous by construction, similarly to what is achieved by the
DCA$^{+}$ method\cite{Staar2013,Staar2014}, but without arbitrary
interpolation schemes.

\subsection{Local components of\emph{ $\mathrm{Im}G$ }and $\mathrm{Im}\Sigma$\label{subsec:Local-components}}

\begin{figure}
\begin{centering}
\includegraphics[width=1\columnwidth]{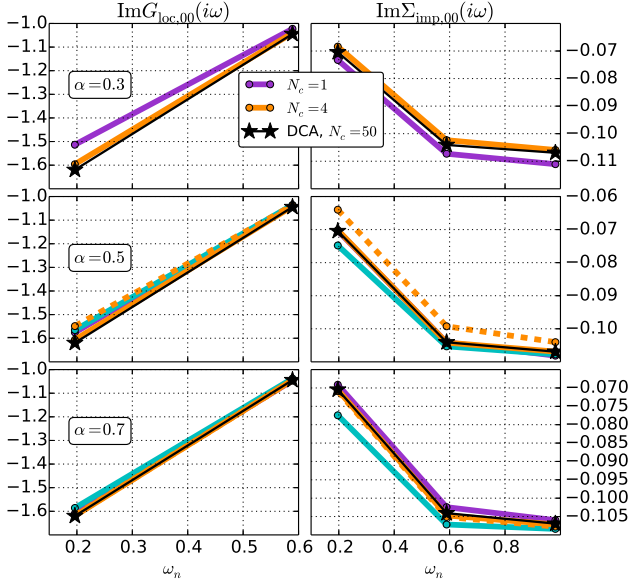} 
\par\end{centering}
\caption{(Point B: $U/D=1$, $\delta=20\%$, $\beta D=16$, $t'=0$, Ising
decoupling). Imaginary part of the local components of $G_{\mathrm{loc}}$
(left column) and $\Sigma_{\mathrm{imp}}$ (right column) for for
$\alpha=0.3$ (top row), $0.5$ (middle row), $0.7$ (bottom row)
and different $N_{c}$. Solid lines: TRILEX. Dashed lines: $GW$+EDMFT
($N_{c}=4$). Black stars: DCA result from Ref. \protect\protect\protect\onlinecite{Leblanc2015},
$N_{c}=50$.\label{fig:local_components}}
\end{figure}

In Fig. \ref{fig:local_components}, we display the local components
$G_{\mathrm{loc}}$ and $\Sigma_{\mathrm{imp}}$ and compare them
to benchmark results obtained with DCA ($N_{c}=50$, Ref. \onlinecite{Leblanc2015}).
The $N_{c}=4$ cluster TRILEX data is the closest to the benchmark
data, irrespective of the value of $\alpha$.

\subsection{Vertex \label{sec:Vertex}}

\subsubsection{Momentum dependence of the vertex}

\begin{figure}[h]
\begin{centering}
\includegraphics[width=0.85\columnwidth]{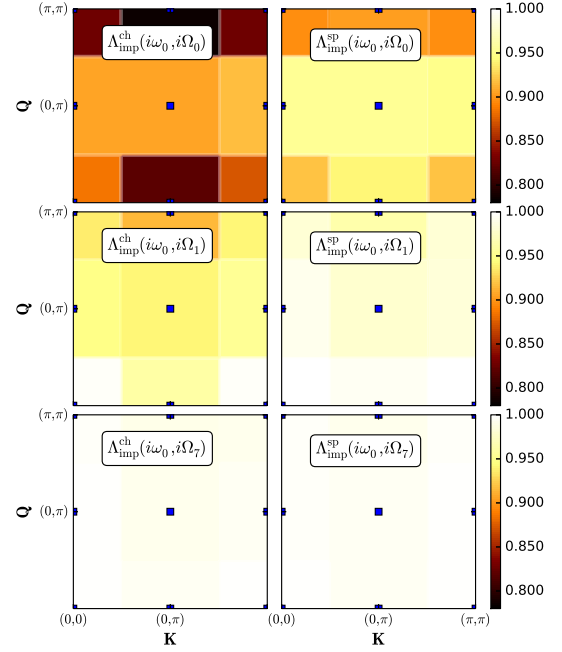} 
\par\end{centering}
\caption{Weak-coupling parameters (Point A, $U/D=0.5$, $\delta=0\%$, $\beta D=16$,
$t'=0$, $\alpha=0.5$, Ising decoupling). Same conventions as Fig.
6 of the main text.\label{fig:Impurity-vertex-weak-coupling}}
\end{figure}

\begin{figure}[H]
\begin{centering}
\includegraphics[width=0.85\columnwidth]{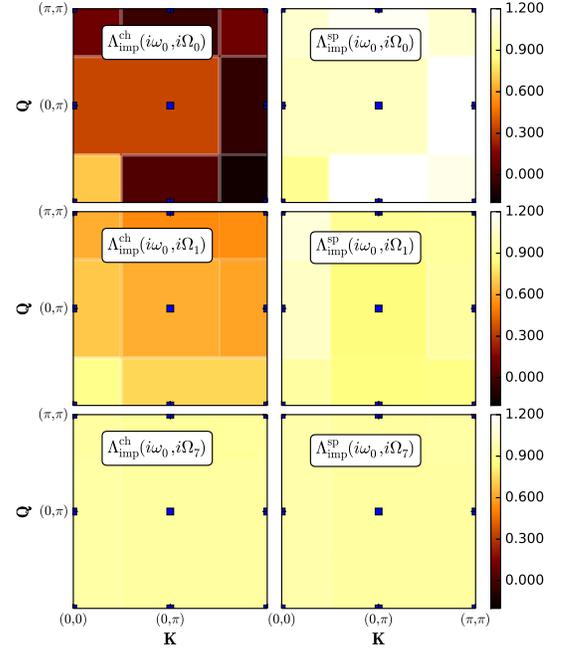} 
\par\end{centering}
\caption{Strong-coupling parameters (Point C, $U/D=1.4$, $\delta=4\%$, $\beta D=8$,
$t'/t=-0.3$, $\alpha=0.5$, Ising decoupling). Same conventions as
Fig. 6 of the main text. \label{fig:Impurity-vertex-strong-coupling}}
\end{figure}

In Figures \ref{fig:Impurity-vertex-weak-coupling} and \ref{fig:Impurity-vertex-strong-coupling},
we show the dependence of the vertex on the cluster momenta $\mathbf{K}$
and $\mathbf{Q}$ for points A and C (point B is shown in the main
text).

\subsubsection{Cluster-site dependence of the vertex\label{sec:Vertex-components}}

\begin{figure*}
\begin{centering}
\includegraphics[width=0.79\textwidth]{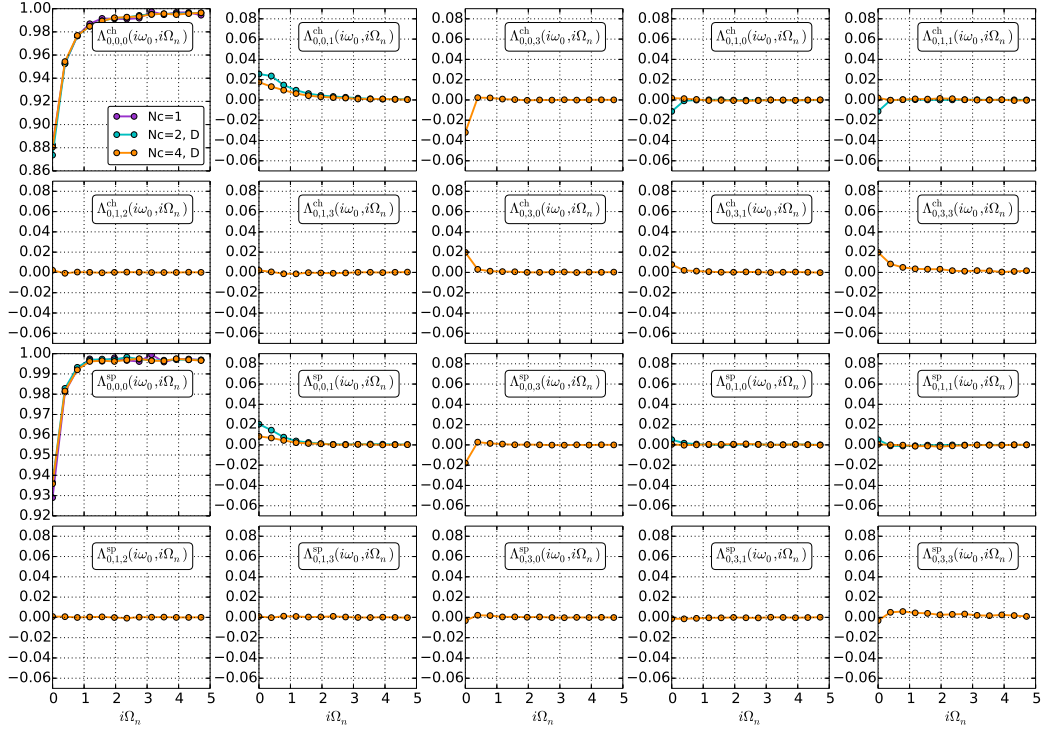} 
\par\end{centering}
\caption{Weak-coupling parameters (Point A, $U/D=0.5$, $\delta=0\%$, $\beta D=16$,
$t'=0$), $\alpha=0.5$, Ising decoupling. Impurity cluster vertex
$\Lambda_{\mathrm{imp}}^{\eta}(i,j,k;i\omega,i\Omega)$ in the charge
(first two rows) and spin (last two rows) channels, at fixed fermionic
Matsubara frequency $\omega_{0}$. See Fig 1 for a definition of the
cluster coordinates $\mathbf{R}_{i},\mathbf{R}_{j}$ and $\mathbf{R}_{k}$
denoted by the indices $i,j,k$. \label{fig:Impurity-cluster-vertex-weak-coupling}}
\end{figure*}

\begin{figure*}
\begin{centering}
\includegraphics[width=0.79\textwidth]{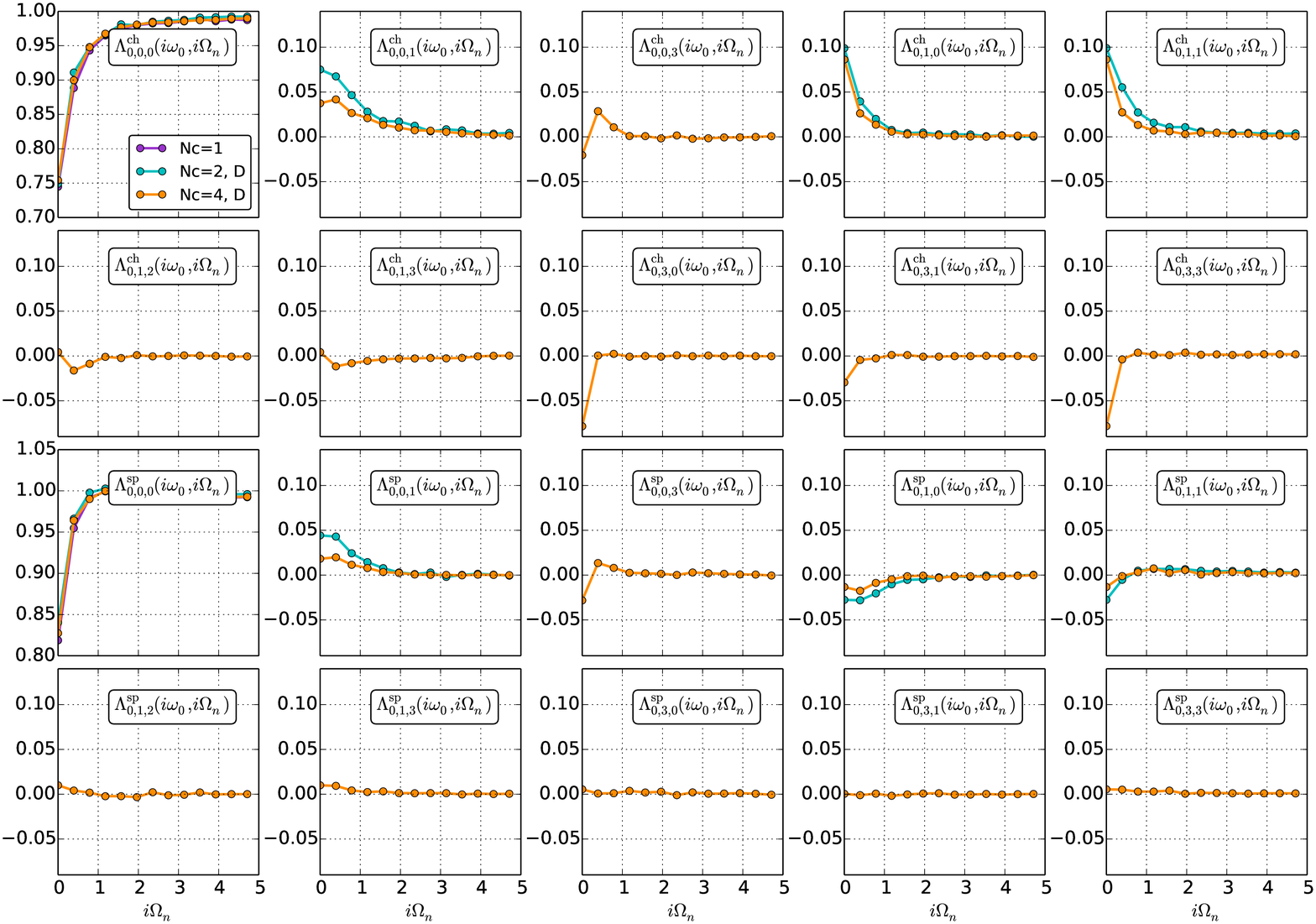} 
\par\end{centering}
\caption{Intermediate-coupling parameters (point B, $U/D=1$, $\delta=20\%$,
$\beta D=16$, $t'=0$), $\alpha=0.5$, Ising decoupling. Impurity
cluster vertex $\Lambda_{\mathrm{imp}}^{\eta}(i,j,k;i\omega,i\Omega)$
in the charge and spin channels, at fixed fermionic Matsubara frequency
$\omega_{0}$. \label{fig:Impurity-cluster-vertex-interm-coupling}}
\end{figure*}

\begin{figure*}
\begin{centering}
\includegraphics[width=0.79\textwidth]{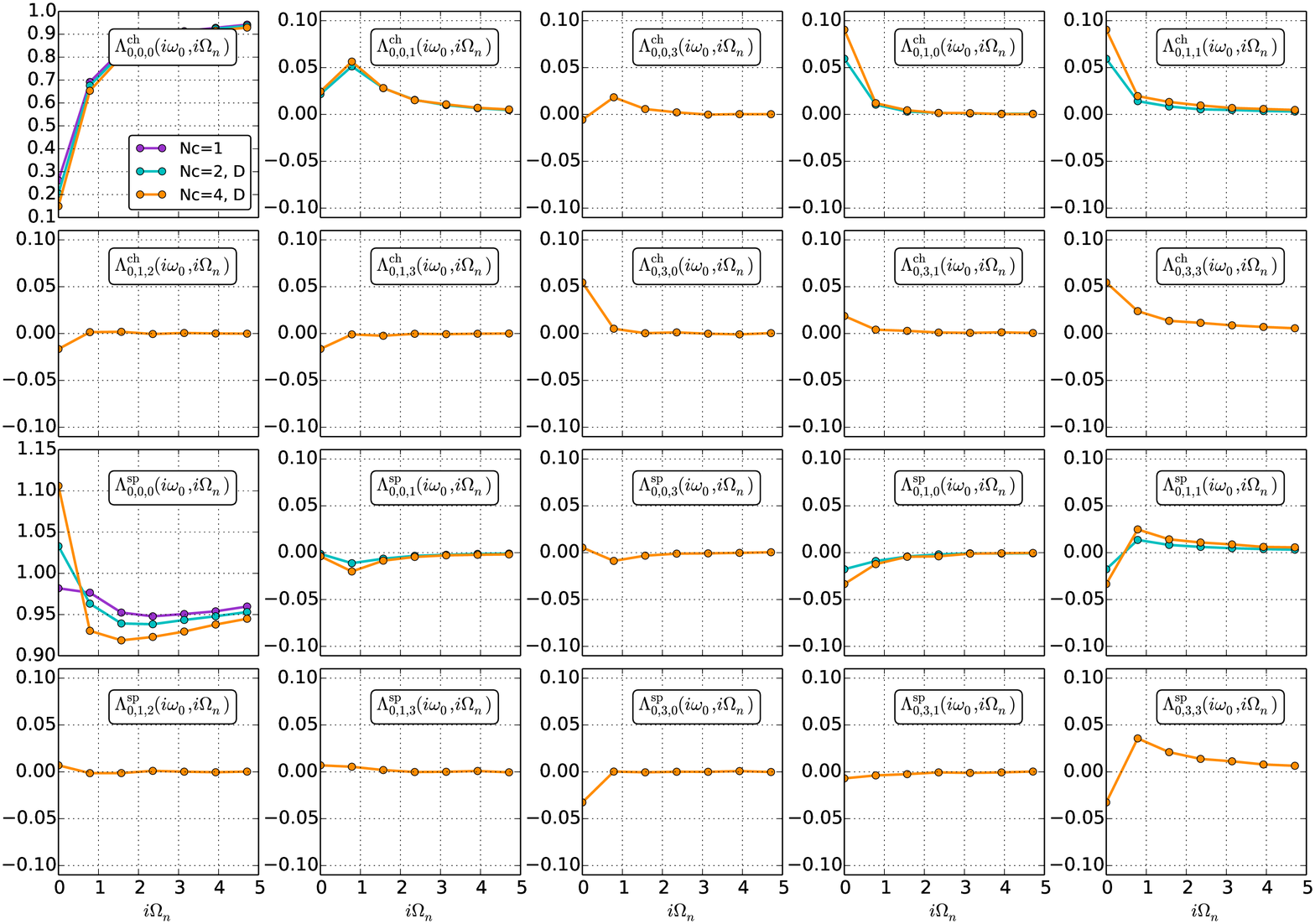} 
\par\end{centering}
\caption{Strong-coupling parameters (Point C, $U/D=1.4$, $\delta=4\%$, $\beta D=8$,
$t'/t=-0.3$), $\alpha=0.5$, Ising decoupling. Impurity cluster vertex
$\Lambda_{\mathrm{imp}}^{\eta}(i,j,k;i\omega,i\Omega)$ in the charge
and spin channels, at fixed fermionic Matsubara frequency $\omega_{0}$.
\label{fig:Impurity-cluster-vertex-strong-coupling}}
\end{figure*}

In Figures \ref{fig:Impurity-cluster-vertex-weak-coupling}, \ref{fig:Impurity-cluster-vertex-interm-coupling}
and \ref{fig:Impurity-cluster-vertex-strong-coupling}, we show all
the inequivalent vertex components $\Lambda_{\mathrm{imp}}(i,j,k;i\omega,i\Omega)$
for the three regimes of parameters (respectively point A, B and C)
studied in the main text. While the largest component is the local
component ($i,j,k=0,0,0$), some nonlocal components are non-negligible.

 \bibliographystyle{apsrev4-1}
\bibliography{cluster_trilex}

\begin{thebibliography}{92}%
\makeatletter
\providecommand \@ifxundefined [1]{%
 \@ifx{#1\undefined}
}%
\providecommand \@ifnum [1]{%
 \ifnum #1\expandafter \@firstoftwo
 \else \expandafter \@secondoftwo
 \fi
}%
\providecommand \@ifx [1]{%
 \ifx #1\expandafter \@firstoftwo
 \else \expandafter \@secondoftwo
 \fi
}%
\providecommand \natexlab [1]{#1}%
\providecommand \enquote  [1]{``#1''}%
\providecommand \bibnamefont  [1]{#1}%
\providecommand \bibfnamefont [1]{#1}%
\providecommand \citenamefont [1]{#1}%
\providecommand \href@noop [0]{\@secondoftwo}%
\providecommand \href [0]{\begingroup \@sanitize@url \@href}%
\providecommand \@href[1]{\@@startlink{#1}\@@href}%
\providecommand \@@href[1]{\endgroup#1\@@endlink}%
\providecommand \@sanitize@url [0]{\catcode `\\12\catcode `\$12\catcode
  `\&12\catcode `\#12\catcode `\^12\catcode `\_12\catcode `\%12\relax}%
\providecommand \@@startlink[1]{}%
\providecommand \@@endlink[0]{}%
\providecommand \url  [0]{\begingroup\@sanitize@url \@url }%
\providecommand \@url [1]{\endgroup\@href {#1}{\urlprefix }}%
\providecommand \urlprefix  [0]{URL }%
\providecommand \Eprint [0]{\href }%
\providecommand \doibase [0]{http://dx.doi.org/}%
\providecommand \selectlanguage [0]{\@gobble}%
\providecommand \bibinfo  [0]{\@secondoftwo}%
\providecommand \bibfield  [0]{\@secondoftwo}%
\providecommand \translation [1]{[#1]}%
\providecommand \BibitemOpen [0]{}%
\providecommand \bibitemStop [0]{}%
\providecommand \bibitemNoStop [0]{.\EOS\space}%
\providecommand \EOS [0]{\spacefactor3000\relax}%
\providecommand \BibitemShut  [1]{\csname bibitem#1\endcsname}%
\let\auto@bib@innerbib\@empty
\bibitem [{\citenamefont {Chubukov}\ \emph {et~al.}(2002)\citenamefont
  {Chubukov}, \citenamefont {Pines},\ and\ \citenamefont
  {Schmalian}}]{Chubukov2002}%
  \BibitemOpen
  \bibfield  {author} {\bibinfo {author} {\bibfnamefont {A.~V.}\ \bibnamefont
  {Chubukov}}, \bibinfo {author} {\bibfnamefont {D.}~\bibnamefont {Pines}}, \
  and\ \bibinfo {author} {\bibfnamefont {J.}~\bibnamefont {Schmalian}},\ }in\
  \href {\doibase 10.1007/978-3-540-73253-2_22} {\emph {\bibinfo {booktitle}
  {Superconductivity}}}\ (\bibinfo  {publisher} {Springer Berlin Heidelberg},\
  \bibinfo {address} {Berlin, Heidelberg},\ \bibinfo {year} {2002})\
  Chap.~\bibinfo {chapter} {22}, pp.\ \bibinfo {pages} {1349--1413}\BibitemShut
  {NoStop}%
\bibitem [{\citenamefont {Onufrieva}\ and\ \citenamefont
  {Pfeuty}(2009)}]{Onufrieva2009}%
  \BibitemOpen
  \bibfield  {author} {\bibinfo {author} {\bibfnamefont {F.}~\bibnamefont
  {Onufrieva}}\ and\ \bibinfo {author} {\bibfnamefont {P.}~\bibnamefont
  {Pfeuty}},\ }\href {\doibase 10.1103/PhysRevLett.102.207003} {\bibfield
  {journal} {\bibinfo  {journal} {Physical Review Letters}\ }\textbf {\bibinfo
  {volume} {102}},\ \bibinfo {pages} {207003} (\bibinfo {year}
  {2009})}\BibitemShut {NoStop}%
\bibitem [{\citenamefont {Metlitski}\ and\ \citenamefont
  {Sachdev}(2010)}]{Metlitski2010}%
  \BibitemOpen
  \bibfield  {author} {\bibinfo {author} {\bibfnamefont {M.~A.}\ \bibnamefont
  {Metlitski}}\ and\ \bibinfo {author} {\bibfnamefont {S.}~\bibnamefont
  {Sachdev}},\ }\href {\doibase 10.1103/PhysRevB.82.075128} {\bibfield
  {journal} {\bibinfo  {journal} {Physical Review B}\ }\textbf {\bibinfo
  {volume} {82}},\ \bibinfo {pages} {075128} (\bibinfo {year}
  {2010})}\BibitemShut {NoStop}%
\bibitem [{\citenamefont {Efetov}\ \emph {et~al.}(2013)\citenamefont {Efetov},
  \citenamefont {Meier},\ and\ \citenamefont {P{\'{e}}pin}}]{Efetov2013}%
  \BibitemOpen
  \bibfield  {author} {\bibinfo {author} {\bibfnamefont {K.~B.}\ \bibnamefont
  {Efetov}}, \bibinfo {author} {\bibfnamefont {H.}~\bibnamefont {Meier}}, \
  and\ \bibinfo {author} {\bibfnamefont {C.}~\bibnamefont {P{\'{e}}pin}},\
  }\href {\doibase 10.1038/nphys2641} {\bibfield  {journal} {\bibinfo
  {journal} {Nature Physics}\ }\textbf {\bibinfo {volume} {9}},\ \bibinfo
  {pages} {442} (\bibinfo {year} {2013})}\BibitemShut {NoStop}%
\bibitem [{\citenamefont {Onufrieva}\ and\ \citenamefont
  {Pfeuty}(2012)}]{Onufrieva2012}%
  \BibitemOpen
  \bibfield  {author} {\bibinfo {author} {\bibfnamefont {F.}~\bibnamefont
  {Onufrieva}}\ and\ \bibinfo {author} {\bibfnamefont {P.}~\bibnamefont
  {Pfeuty}},\ }\href {\doibase 10.1103/PhysRevLett.109.257001} {\bibfield
  {journal} {\bibinfo  {journal} {Physical Review Letters}\ }\textbf {\bibinfo
  {volume} {109}},\ \bibinfo {pages} {257001} (\bibinfo {year}
  {2012})}\BibitemShut {NoStop}%
\bibitem [{\citenamefont {Scalapino}(2012)}]{Scalapino2012}%
  \BibitemOpen
  \bibfield  {author} {\bibinfo {author} {\bibfnamefont {D.~J.}\ \bibnamefont
  {Scalapino}},\ }\href {\doibase 10.1103/RevModPhys.84.1383} {\bibfield
  {journal} {\bibinfo  {journal} {Reviews of Modern Physics}\ }\textbf
  {\bibinfo {volume} {84}},\ \bibinfo {pages} {1383} (\bibinfo {year}
  {2012})}\BibitemShut {NoStop}%
\bibitem [{\citenamefont {Wang}\ and\ \citenamefont
  {Chubukov}(2014)}]{Wang2014}%
  \BibitemOpen
  \bibfield  {author} {\bibinfo {author} {\bibfnamefont {Y.}~\bibnamefont
  {Wang}}\ and\ \bibinfo {author} {\bibfnamefont {A.}~\bibnamefont
  {Chubukov}},\ }\href {\doibase 10.1103/PhysRevB.90.035149} {\bibfield
  {journal} {\bibinfo  {journal} {Physical Review B}\ }\textbf {\bibinfo
  {volume} {90}},\ \bibinfo {pages} {035149} (\bibinfo {year}
  {2014})}\BibitemShut {NoStop}%
\bibitem [{\citenamefont {Wang}\ \emph {et~al.}(2016)\citenamefont {Wang},
  \citenamefont {Abanov}, \citenamefont {Altshuler}, \citenamefont
  {Yuzbashyan},\ and\ \citenamefont {Chubukov}}]{Wang2016}%
  \BibitemOpen
  \bibfield  {author} {\bibinfo {author} {\bibfnamefont {Y.}~\bibnamefont
  {Wang}}, \bibinfo {author} {\bibfnamefont {A.}~\bibnamefont {Abanov}},
  \bibinfo {author} {\bibfnamefont {B.~L.}\ \bibnamefont {Altshuler}}, \bibinfo
  {author} {\bibfnamefont {E.~A.}\ \bibnamefont {Yuzbashyan}}, \ and\ \bibinfo
  {author} {\bibfnamefont {A.~V.}\ \bibnamefont {Chubukov}},\ }\href {\doibase
  10.1103/PhysRevLett.117.157001} {\bibfield  {journal} {\bibinfo  {journal}
  {Physical Review Letters}\ }\textbf {\bibinfo {volume} {117}},\ \bibinfo
  {pages} {157001} (\bibinfo {year} {2016})}\BibitemShut {NoStop}%
\bibitem [{\citenamefont {Bickers}\ and\ \citenamefont
  {Scalapino}(1989)}]{Bickers1989}%
  \BibitemOpen
  \bibfield  {author} {\bibinfo {author} {\bibfnamefont {N.}~\bibnamefont
  {Bickers}}\ and\ \bibinfo {author} {\bibfnamefont {D.}~\bibnamefont
  {Scalapino}},\ }\href {http://www.sciencedirect.com/science/article/pii/
  000349168990359X} {\bibfield  {journal} {\bibinfo  {journal} {Annals of
  Physics}\ }\textbf {\bibinfo {volume} {206251}},\ \bibinfo {pages} {206}
  (\bibinfo {year} {1989})}\BibitemShut {NoStop}%
\bibitem [{\citenamefont {Hedin}(1965)}]{Hedin1965}%
  \BibitemOpen
  \bibfield  {author} {\bibinfo {author} {\bibfnamefont {L.}~\bibnamefont
  {Hedin}},\ }\href {http://journals.aps.org/pr/abstract/10.1103/
  PhysRev.139.A796} {\bibfield  {journal} {\bibinfo  {journal} {Physical
  Review}\ }\textbf {\bibinfo {volume} {139}},\ \bibinfo {pages} {796}
  (\bibinfo {year} {1965})}\BibitemShut {NoStop}%
\bibitem [{\citenamefont {Hedin}(1999)}]{Hedin1999}%
  \BibitemOpen
  \bibfield  {author} {\bibinfo {author} {\bibfnamefont {L.}~\bibnamefont
  {Hedin}},\ }\href {http://iopscience.iop.org/0953-8984/11/42/201} {\bibfield
  {journal} {\bibinfo  {journal} {Journal of Physics: Condensed Matter}\
  }\textbf {\bibinfo {volume} {489}} (\bibinfo {year} {1999})}\BibitemShut
  {NoStop}%
\bibitem [{\citenamefont {Eliashberg}(1960)}]{Eliashberg1960}%
  \BibitemOpen
  \bibfield  {author} {\bibinfo {author} {\bibfnamefont {G.~M.}\ \bibnamefont
  {Eliashberg}},\ }\href@noop {} {\bibfield  {journal} {\bibinfo  {journal}
  {Sov. Phys. JETP}\ }\textbf {\bibinfo {volume} {11}},\ \bibinfo {pages} {696}
  (\bibinfo {year} {1960})}\BibitemShut {NoStop}%
\bibitem [{\citenamefont {Jaeckel}\ and\ \citenamefont
  {Wetterich}(2003)}]{Jaeckel2003}%
  \BibitemOpen
  \bibfield  {author} {\bibinfo {author} {\bibfnamefont {J.}~\bibnamefont
  {Jaeckel}}\ and\ \bibinfo {author} {\bibfnamefont {C.}~\bibnamefont
  {Wetterich}},\ }\href {\doibase 10.1103/PhysRevD.68.025020} {\bibfield
  {journal} {\bibinfo  {journal} {Physical Review D}\ }\textbf {\bibinfo
  {volume} {68}},\ \bibinfo {pages} {025020} (\bibinfo {year}
  {2003})}\BibitemShut {NoStop}%
\bibitem [{\citenamefont {Baier}\ \emph {et~al.}(2004)\citenamefont {Baier},
  \citenamefont {Bick},\ and\ \citenamefont {Wetterich}}]{Baier2004}%
  \BibitemOpen
  \bibfield  {author} {\bibinfo {author} {\bibfnamefont {T.}~\bibnamefont
  {Baier}}, \bibinfo {author} {\bibfnamefont {E.}~\bibnamefont {Bick}}, \ and\
  \bibinfo {author} {\bibfnamefont {C.}~\bibnamefont {Wetterich}},\ }\href
  {\doibase 10.1103/PhysRevB.70.125111} {\bibfield  {journal} {\bibinfo
  {journal} {Physical Review B}\ }\textbf {\bibinfo {volume} {70}},\ \bibinfo
  {pages} {125111} (\bibinfo {year} {2004})}\BibitemShut {NoStop}%
\bibitem [{\citenamefont {Bartosch}\ \emph {et~al.}(2009)\citenamefont
  {Bartosch}, \citenamefont {Freire}, \citenamefont {Cardenas},\ and\
  \citenamefont {Kopietz}}]{Bartosch2009}%
  \BibitemOpen
  \bibfield  {author} {\bibinfo {author} {\bibfnamefont {L.}~\bibnamefont
  {Bartosch}}, \bibinfo {author} {\bibfnamefont {H.}~\bibnamefont {Freire}},
  \bibinfo {author} {\bibfnamefont {J.~J.~R.}\ \bibnamefont {Cardenas}}, \ and\
  \bibinfo {author} {\bibfnamefont {P.}~\bibnamefont {Kopietz}},\ }\href
  {\doibase 10.1088/0953-8984/21/30/305602} {\bibfield  {journal} {\bibinfo
  {journal} {Journal of Physics: Condensed Matter}\ }\textbf {\bibinfo {volume}
  {21}},\ \bibinfo {pages} {305602} (\bibinfo {year} {2009})}\BibitemShut
  {NoStop}%
\bibitem [{\citenamefont {Borejsza}\ and\ \citenamefont
  {Dupuis}(2003)}]{Borejsza2003}%
  \BibitemOpen
  \bibfield  {author} {\bibinfo {author} {\bibfnamefont {K.}~\bibnamefont
  {Borejsza}}\ and\ \bibinfo {author} {\bibfnamefont {N.}~\bibnamefont
  {Dupuis}},\ }\href {\doibase 10.1209/epl/i2003-00584-7} {\bibfield  {journal}
  {\bibinfo  {journal} {Europhysics Letters (EPL)}\ }\textbf {\bibinfo {volume}
  {63}},\ \bibinfo {pages} {722} (\bibinfo {year} {2003})}\BibitemShut
  {NoStop}%
\bibitem [{\citenamefont {Borejsza}\ and\ \citenamefont
  {Dupuis}(2004)}]{Borejsza2004}%
  \BibitemOpen
  \bibfield  {author} {\bibinfo {author} {\bibfnamefont {K.}~\bibnamefont
  {Borejsza}}\ and\ \bibinfo {author} {\bibfnamefont {N.}~\bibnamefont
  {Dupuis}},\ }\href {\doibase 10.1103/PhysRevB.69.085119} {\bibfield
  {journal} {\bibinfo  {journal} {Physical Review B}\ }\textbf {\bibinfo
  {volume} {69}},\ \bibinfo {pages} {085119} (\bibinfo {year}
  {2004})}\BibitemShut {NoStop}%
\bibitem [{\citenamefont {Dupuis}(2002)}]{Dupuis2002}%
  \BibitemOpen
  \bibfield  {author} {\bibinfo {author} {\bibfnamefont {N.}~\bibnamefont
  {Dupuis}},\ }\href {\doibase 10.1103/PhysRevB.65.245118} {\bibfield
  {journal} {\bibinfo  {journal} {Physical Review B}\ }\textbf {\bibinfo
  {volume} {65}},\ \bibinfo {pages} {245118} (\bibinfo {year}
  {2002})}\BibitemShut {NoStop}%
\bibitem [{\citenamefont {Anderson}(1987)}]{Anderson1987}%
  \BibitemOpen
  \bibfield  {author} {\bibinfo {author} {\bibfnamefont {P.~W.}\ \bibnamefont
  {Anderson}},\ }\href {http://www.sciencemag.org/content/235/4793/1196.short}
  {\bibfield  {journal} {\bibinfo  {journal} {Science}\ }\textbf {\bibinfo
  {volume} {235}},\ \bibinfo {pages} {1196} (\bibinfo {year}
  {1987})}\BibitemShut {NoStop}%
\bibitem [{\citenamefont {Georges}\ \emph {et~al.}(1996)\citenamefont
  {Georges}, \citenamefont {Kotliar}, \citenamefont {Krauth},\ and\
  \citenamefont {Rozenberg}}]{Georges1996}%
  \BibitemOpen
  \bibfield  {author} {\bibinfo {author} {\bibfnamefont {A.}~\bibnamefont
  {Georges}}, \bibinfo {author} {\bibfnamefont {G.}~\bibnamefont {Kotliar}},
  \bibinfo {author} {\bibfnamefont {W.}~\bibnamefont {Krauth}}, \ and\ \bibinfo
  {author} {\bibfnamefont {M.~J.}\ \bibnamefont {Rozenberg}},\ }\href {\doibase
  10.1103/RevModPhys.68.13} {\bibfield  {journal} {\bibinfo  {journal} {Reviews
  of Modern Physics}\ }\textbf {\bibinfo {volume} {68}},\ \bibinfo {pages} {13}
  (\bibinfo {year} {1996})}\BibitemShut {NoStop}%
\bibitem [{\citenamefont {Hettler}\ \emph {et~al.}(1998)\citenamefont
  {Hettler}, \citenamefont {Tahvildar-Zadeh}, \citenamefont {Jarrell},
  \citenamefont {Pruschke},\ and\ \citenamefont {Krishnamurthy}}]{Hettler1998}%
  \BibitemOpen
  \bibfield  {author} {\bibinfo {author} {\bibfnamefont {M.~H.}\ \bibnamefont
  {Hettler}}, \bibinfo {author} {\bibfnamefont {A.~N.}\ \bibnamefont
  {Tahvildar-Zadeh}}, \bibinfo {author} {\bibfnamefont {M.}~\bibnamefont
  {Jarrell}}, \bibinfo {author} {\bibfnamefont {T.}~\bibnamefont {Pruschke}}, \
  and\ \bibinfo {author} {\bibfnamefont {H.~R.}\ \bibnamefont
  {Krishnamurthy}},\ }\href {\doibase 10.1103/PhysRevB.58.R7475} {\bibfield
  {journal} {\bibinfo  {journal} {Physical Review B}\ }\textbf {\bibinfo
  {volume} {58}},\ \bibinfo {pages} {R7475} (\bibinfo {year}
  {1998})}\BibitemShut {NoStop}%
\bibitem [{\citenamefont {Hettler}\ \emph {et~al.}(1999)\citenamefont
  {Hettler}, \citenamefont {Mukherjee}, \citenamefont {Jarrell},\ and\
  \citenamefont {Krishnamurthy}}]{Hettler1999}%
  \BibitemOpen
  \bibfield  {author} {\bibinfo {author} {\bibfnamefont {M.~H.}\ \bibnamefont
  {Hettler}}, \bibinfo {author} {\bibfnamefont {M.}~\bibnamefont {Mukherjee}},
  \bibinfo {author} {\bibfnamefont {M.}~\bibnamefont {Jarrell}}, \ and\
  \bibinfo {author} {\bibfnamefont {H.~R.}\ \bibnamefont {Krishnamurthy}},\
  }\href {\doibase 10.1103/PhysRevB.61.12739} {\bibfield  {journal} {\bibinfo
  {journal} {Physical Review B}\ }\textbf {\bibinfo {volume} {61}},\ \bibinfo
  {pages} {12739} (\bibinfo {year} {1999})}\BibitemShut {NoStop}%
\bibitem [{\citenamefont {Lichtenstein}\ and\ \citenamefont
  {Katsnelson}(2000)}]{Lichtenstein2000}%
  \BibitemOpen
  \bibfield  {author} {\bibinfo {author} {\bibfnamefont {A.~I.}\ \bibnamefont
  {Lichtenstein}}\ and\ \bibinfo {author} {\bibfnamefont {M.~I.}\ \bibnamefont
  {Katsnelson}},\ }\href {\doibase 10.1103/PhysRevB.62.R9283} {\bibfield
  {journal} {\bibinfo  {journal} {Physical Review B}\ }\textbf {\bibinfo
  {volume} {62}},\ \bibinfo {pages} {R9283} (\bibinfo {year}
  {2000})}\BibitemShut {NoStop}%
\bibitem [{\citenamefont {Kotliar}\ \emph {et~al.}(2001)\citenamefont
  {Kotliar}, \citenamefont {Savrasov}, \citenamefont {P{\'{a}}lsson},\ and\
  \citenamefont {Biroli}}]{Kotliar2001}%
  \BibitemOpen
  \bibfield  {author} {\bibinfo {author} {\bibfnamefont {G.}~\bibnamefont
  {Kotliar}}, \bibinfo {author} {\bibfnamefont {S.}~\bibnamefont {Savrasov}},
  \bibinfo {author} {\bibfnamefont {G.}~\bibnamefont {P{\'{a}}lsson}}, \ and\
  \bibinfo {author} {\bibfnamefont {G.}~\bibnamefont {Biroli}},\ }\href
  {\doibase 10.1103/PhysRevLett.87.186401} {\bibfield  {journal} {\bibinfo
  {journal} {Physical Review Letters}\ }\textbf {\bibinfo {volume} {87}},\
  \bibinfo {pages} {186401} (\bibinfo {year} {2001})}\BibitemShut {NoStop}%
\bibitem [{\citenamefont {Maier}\ \emph
  {et~al.}(2005{\natexlab{a}})\citenamefont {Maier}, \citenamefont {Jarrell},
  \citenamefont {Pruschke},\ and\ \citenamefont {Hettler}}]{Maier2005a}%
  \BibitemOpen
  \bibfield  {author} {\bibinfo {author} {\bibfnamefont {T.~A.}\ \bibnamefont
  {Maier}}, \bibinfo {author} {\bibfnamefont {M.}~\bibnamefont {Jarrell}},
  \bibinfo {author} {\bibfnamefont {T.}~\bibnamefont {Pruschke}}, \ and\
  \bibinfo {author} {\bibfnamefont {M.~H.}\ \bibnamefont {Hettler}},\ }\href
  {\doibase 10.1103/RevModPhys.77.1027} {\bibfield  {journal} {\bibinfo
  {journal} {Reviews of Modern Physics}\ }\textbf {\bibinfo {volume} {77}},\
  \bibinfo {pages} {1027} (\bibinfo {year} {2005}{\natexlab{a}})}\BibitemShut
  {NoStop}%
\bibitem [{\citenamefont {Kyung}\ \emph {et~al.}(2009)\citenamefont {Kyung},
  \citenamefont {S{\'{e}}n{\'{e}}chal},\ and\ \citenamefont
  {Tremblay}}]{Kyung2009}%
  \BibitemOpen
  \bibfield  {author} {\bibinfo {author} {\bibfnamefont {B.}~\bibnamefont
  {Kyung}}, \bibinfo {author} {\bibfnamefont {D.}~\bibnamefont
  {S{\'{e}}n{\'{e}}chal}}, \ and\ \bibinfo {author} {\bibfnamefont {A.-M.~S.}\
  \bibnamefont {Tremblay}},\ }\href {\doibase 10.1103/PhysRevB.80.205109}
  {\bibfield  {journal} {\bibinfo  {journal} {Physical Review B}\ }\textbf
  {\bibinfo {volume} {80}},\ \bibinfo {pages} {205109} (\bibinfo {year}
  {2009})}\BibitemShut {NoStop}%
\bibitem [{\citenamefont {Sordi}\ \emph
  {et~al.}(2012{\natexlab{a}})\citenamefont {Sordi}, \citenamefont
  {S{\'{e}}mon}, \citenamefont {Haule},\ and\ \citenamefont
  {Tremblay}}]{Sordi2012}%
  \BibitemOpen
  \bibfield  {author} {\bibinfo {author} {\bibfnamefont {G.}~\bibnamefont
  {Sordi}}, \bibinfo {author} {\bibfnamefont {P.}~\bibnamefont {S{\'{e}}mon}},
  \bibinfo {author} {\bibfnamefont {K.}~\bibnamefont {Haule}}, \ and\ \bibinfo
  {author} {\bibfnamefont {A.-M.~S.}\ \bibnamefont {Tremblay}},\ }\href
  {\doibase 10.1103/PhysRevLett.108.216401} {\bibfield  {journal} {\bibinfo
  {journal} {Physical Review Letters}\ }\textbf {\bibinfo {volume} {108}},\
  \bibinfo {pages} {216401} (\bibinfo {year} {2012}{\natexlab{a}})}\BibitemShut
  {NoStop}%
\bibitem [{\citenamefont {Civelli}\ \emph {et~al.}(2008)\citenamefont
  {Civelli}, \citenamefont {Capone}, \citenamefont {Georges}, \citenamefont
  {Haule}, \citenamefont {Parcollet}, \citenamefont {Stanescu},\ and\
  \citenamefont {Kotliar}}]{Civelli2008}%
  \BibitemOpen
  \bibfield  {author} {\bibinfo {author} {\bibfnamefont {M.}~\bibnamefont
  {Civelli}}, \bibinfo {author} {\bibfnamefont {M.}~\bibnamefont {Capone}},
  \bibinfo {author} {\bibfnamefont {A.}~\bibnamefont {Georges}}, \bibinfo
  {author} {\bibfnamefont {K.}~\bibnamefont {Haule}}, \bibinfo {author}
  {\bibfnamefont {O.}~\bibnamefont {Parcollet}}, \bibinfo {author}
  {\bibfnamefont {T.~D.}\ \bibnamefont {Stanescu}}, \ and\ \bibinfo {author}
  {\bibfnamefont {G.}~\bibnamefont {Kotliar}},\ }\href {\doibase
  10.1103/PhysRevLett.100.046402} {\bibfield  {journal} {\bibinfo  {journal}
  {Physical Review Letters}\ }\textbf {\bibinfo {volume} {100}},\ \bibinfo
  {pages} {046402} (\bibinfo {year} {2008})}\BibitemShut {NoStop}%
\bibitem [{\citenamefont {Ferrero}\ \emph {et~al.}(2010)\citenamefont
  {Ferrero}, \citenamefont {Parcollet}, \citenamefont {Georges}, \citenamefont
  {Kotliar},\ and\ \citenamefont {Basov}}]{Ferrero2010}%
  \BibitemOpen
  \bibfield  {author} {\bibinfo {author} {\bibfnamefont {M.}~\bibnamefont
  {Ferrero}}, \bibinfo {author} {\bibfnamefont {O.}~\bibnamefont {Parcollet}},
  \bibinfo {author} {\bibfnamefont {a.}~\bibnamefont {Georges}}, \bibinfo
  {author} {\bibfnamefont {G.}~\bibnamefont {Kotliar}}, \ and\ \bibinfo
  {author} {\bibfnamefont {D.~N.}\ \bibnamefont {Basov}},\ }\href {\doibase
  10.1103/PhysRevB.82.054502} {\bibfield  {journal} {\bibinfo  {journal}
  {Physical Review B}\ }\textbf {\bibinfo {volume} {82}},\ \bibinfo {pages}
  {054502} (\bibinfo {year} {2010})}\BibitemShut {NoStop}%
\bibitem [{\citenamefont {Gull}\ \emph {et~al.}(2013)\citenamefont {Gull},
  \citenamefont {Parcollet},\ and\ \citenamefont {Millis}}]{Gull2013}%
  \BibitemOpen
  \bibfield  {author} {\bibinfo {author} {\bibfnamefont {E.}~\bibnamefont
  {Gull}}, \bibinfo {author} {\bibfnamefont {O.}~\bibnamefont {Parcollet}}, \
  and\ \bibinfo {author} {\bibfnamefont {A.~J.}\ \bibnamefont {Millis}},\
  }\href {\doibase 10.1103/PhysRevLett.110.216405} {\bibfield  {journal}
  {\bibinfo  {journal} {Physical Review Letters}\ }\textbf {\bibinfo {volume}
  {110}},\ \bibinfo {pages} {216405} (\bibinfo {year} {2013})}\BibitemShut
  {NoStop}%
\bibitem [{\citenamefont {Macridin}\ \emph {et~al.}(2004)\citenamefont
  {Macridin}, \citenamefont {Jarrell},\ and\ \citenamefont
  {Maier}}]{Macridin2004}%
  \BibitemOpen
  \bibfield  {author} {\bibinfo {author} {\bibfnamefont {A.}~\bibnamefont
  {Macridin}}, \bibinfo {author} {\bibfnamefont {M.}~\bibnamefont {Jarrell}}, \
  and\ \bibinfo {author} {\bibfnamefont {T.~A.}\ \bibnamefont {Maier}},\ }\href
  {\doibase 10.1103/PhysRevB.70.113105} {\bibfield  {journal} {\bibinfo
  {journal} {Physical Review B}\ }\textbf {\bibinfo {volume} {70}},\ \bibinfo
  {pages} {113105} (\bibinfo {year} {2004})}\BibitemShut {NoStop}%
\bibitem [{\citenamefont {Maier}\ \emph {et~al.}(2004)\citenamefont {Maier},
  \citenamefont {Jarrell}, \citenamefont {Macridin},\ and\ \citenamefont
  {Slezak}}]{Maier2004}%
  \BibitemOpen
  \bibfield  {author} {\bibinfo {author} {\bibfnamefont {T.~A.}\ \bibnamefont
  {Maier}}, \bibinfo {author} {\bibfnamefont {M.}~\bibnamefont {Jarrell}},
  \bibinfo {author} {\bibfnamefont {A.}~\bibnamefont {Macridin}}, \ and\
  \bibinfo {author} {\bibfnamefont {C.}~\bibnamefont {Slezak}},\ }\href
  {\doibase 10.1103/PhysRevLett.92.027005} {\bibfield  {journal} {\bibinfo
  {journal} {Physical Review Letters}\ }\textbf {\bibinfo {volume} {92}},\
  \bibinfo {pages} {027005} (\bibinfo {year} {2004})}\BibitemShut {NoStop}%
\bibitem [{\citenamefont {Maier}\ \emph
  {et~al.}(2005{\natexlab{b}})\citenamefont {Maier}, \citenamefont {Jarrell},
  \citenamefont {Schulthess}, \citenamefont {Kent},\ and\ \citenamefont
  {White}}]{Maier2005}%
  \BibitemOpen
  \bibfield  {author} {\bibinfo {author} {\bibfnamefont {T.~A.}\ \bibnamefont
  {Maier}}, \bibinfo {author} {\bibfnamefont {M.}~\bibnamefont {Jarrell}},
  \bibinfo {author} {\bibfnamefont {T.~C.}\ \bibnamefont {Schulthess}},
  \bibinfo {author} {\bibfnamefont {P.~R.~C.}\ \bibnamefont {Kent}}, \ and\
  \bibinfo {author} {\bibfnamefont {J.~B.}\ \bibnamefont {White}},\ }\href
  {\doibase 10.1103/PhysRevLett.95.237001} {\bibfield  {journal} {\bibinfo
  {journal} {Physical Review Letters}\ }\textbf {\bibinfo {volume} {95}},\
  \bibinfo {pages} {237001} (\bibinfo {year} {2005}{\natexlab{b}})}\BibitemShut
  {NoStop}%
\bibitem [{\citenamefont {Maier}\ \emph {et~al.}(2006)\citenamefont {Maier},
  \citenamefont {Jarrell},\ and\ \citenamefont {Scalapino}}]{Maier2006}%
  \BibitemOpen
  \bibfield  {author} {\bibinfo {author} {\bibfnamefont {T.}~\bibnamefont
  {Maier}}, \bibinfo {author} {\bibfnamefont {M.}~\bibnamefont {Jarrell}}, \
  and\ \bibinfo {author} {\bibfnamefont {D.}~\bibnamefont {Scalapino}},\ }\href
  {\doibase 10.1103/PhysRevLett.96.047005} {\bibfield  {journal} {\bibinfo
  {journal} {Physical Review Letters}\ }\textbf {\bibinfo {volume} {96}},\
  \bibinfo {pages} {047005} (\bibinfo {year} {2006})}\BibitemShut {NoStop}%
\bibitem [{\citenamefont {Gull}\ \emph {et~al.}(2010)\citenamefont {Gull},
  \citenamefont {Ferrero}, \citenamefont {Parcollet}, \citenamefont {Georges},\
  and\ \citenamefont {Millis}}]{Gull2010}%
  \BibitemOpen
  \bibfield  {author} {\bibinfo {author} {\bibfnamefont {E.}~\bibnamefont
  {Gull}}, \bibinfo {author} {\bibfnamefont {M.}~\bibnamefont {Ferrero}},
  \bibinfo {author} {\bibfnamefont {O.}~\bibnamefont {Parcollet}}, \bibinfo
  {author} {\bibfnamefont {A.}~\bibnamefont {Georges}}, \ and\ \bibinfo
  {author} {\bibfnamefont {A.~J.}\ \bibnamefont {Millis}},\ }\href {\doibase
  10.1103/PhysRevB.82.155101} {\bibfield  {journal} {\bibinfo  {journal}
  {Physical Review B}\ }\textbf {\bibinfo {volume} {82}},\ \bibinfo {pages}
  {155101} (\bibinfo {year} {2010})}\BibitemShut {NoStop}%
\bibitem [{\citenamefont {Yang}\ \emph
  {et~al.}(2011{\natexlab{a}})\citenamefont {Yang}, \citenamefont {Fotso},
  \citenamefont {Su}, \citenamefont {Galanakis}, \citenamefont {Khatami},
  \citenamefont {She}, \citenamefont {Moreno}, \citenamefont {Zaanen},\ and\
  \citenamefont {Jarrell}}]{Yang2011}%
  \BibitemOpen
  \bibfield  {author} {\bibinfo {author} {\bibfnamefont {S.~X.}\ \bibnamefont
  {Yang}}, \bibinfo {author} {\bibfnamefont {H.}~\bibnamefont {Fotso}},
  \bibinfo {author} {\bibfnamefont {S.~Q.}\ \bibnamefont {Su}}, \bibinfo
  {author} {\bibfnamefont {D.}~\bibnamefont {Galanakis}}, \bibinfo {author}
  {\bibfnamefont {E.}~\bibnamefont {Khatami}}, \bibinfo {author} {\bibfnamefont
  {J.~H.}\ \bibnamefont {She}}, \bibinfo {author} {\bibfnamefont
  {J.}~\bibnamefont {Moreno}}, \bibinfo {author} {\bibfnamefont
  {J.}~\bibnamefont {Zaanen}}, \ and\ \bibinfo {author} {\bibfnamefont
  {M.}~\bibnamefont {Jarrell}},\ }\href {\doibase
  10.1103/PhysRevLett.106.047004} {\bibfield  {journal} {\bibinfo  {journal}
  {Physical Review Letters}\ }\textbf {\bibinfo {volume} {106}},\ \bibinfo
  {pages} {047004} (\bibinfo {year} {2011}{\natexlab{a}})}\BibitemShut
  {NoStop}%
\bibitem [{\citenamefont {Macridin}\ and\ \citenamefont
  {Jarrell}(2008)}]{Macridin2008}%
  \BibitemOpen
  \bibfield  {author} {\bibinfo {author} {\bibfnamefont {A.}~\bibnamefont
  {Macridin}}\ and\ \bibinfo {author} {\bibfnamefont {M.}~\bibnamefont
  {Jarrell}},\ }\href {\doibase 10.1103/PhysRevB.78.241101} {\bibfield
  {journal} {\bibinfo  {journal} {Physical Review B}\ }\textbf {\bibinfo
  {volume} {78}},\ \bibinfo {pages} {241101(R)} (\bibinfo {year}
  {2008})}\BibitemShut {NoStop}%
\bibitem [{\citenamefont {Macridin}\ \emph {et~al.}(2006)\citenamefont
  {Macridin}, \citenamefont {Jarrell}, \citenamefont {Maier}, \citenamefont
  {Kent},\ and\ \citenamefont {D'Azevedo}}]{Macridin2006}%
  \BibitemOpen
  \bibfield  {author} {\bibinfo {author} {\bibfnamefont {A.}~\bibnamefont
  {Macridin}}, \bibinfo {author} {\bibfnamefont {M.}~\bibnamefont {Jarrell}},
  \bibinfo {author} {\bibfnamefont {T.}~\bibnamefont {Maier}}, \bibinfo
  {author} {\bibfnamefont {P.~R.~C.}\ \bibnamefont {Kent}}, \ and\ \bibinfo
  {author} {\bibfnamefont {E.}~\bibnamefont {D'Azevedo}},\ }\href {\doibase
  10.1103/PhysRevLett.97.036401} {\bibfield  {journal} {\bibinfo  {journal}
  {Physical Review Letters}\ }\textbf {\bibinfo {volume} {97}},\ \bibinfo
  {pages} {036401} (\bibinfo {year} {2006})}\BibitemShut {NoStop}%
\bibitem [{\citenamefont {Jarrell}\ \emph {et~al.}(2001)\citenamefont
  {Jarrell}, \citenamefont {Maier}, \citenamefont {Huscroft},\ and\
  \citenamefont {Moukouri}}]{Jarrell2001}%
  \BibitemOpen
  \bibfield  {author} {\bibinfo {author} {\bibfnamefont {M.}~\bibnamefont
  {Jarrell}}, \bibinfo {author} {\bibfnamefont {T.~A.}\ \bibnamefont {Maier}},
  \bibinfo {author} {\bibfnamefont {C.}~\bibnamefont {Huscroft}}, \ and\
  \bibinfo {author} {\bibfnamefont {S.}~\bibnamefont {Moukouri}},\ }\href
  {\doibase 10.1103/PhysRevB.64.195130} {\bibfield  {journal} {\bibinfo
  {journal} {Physical Review B}\ }\textbf {\bibinfo {volume} {64}},\ \bibinfo
  {pages} {195130} (\bibinfo {year} {2001})}\BibitemShut {NoStop}%
\bibitem [{\citenamefont {Parcollet}\ \emph {et~al.}(2004)\citenamefont
  {Parcollet}, \citenamefont {Biroli},\ and\ \citenamefont
  {Kotliar}}]{Parcollet2004}%
  \BibitemOpen
  \bibfield  {author} {\bibinfo {author} {\bibfnamefont {O.}~\bibnamefont
  {Parcollet}}, \bibinfo {author} {\bibfnamefont {G.}~\bibnamefont {Biroli}}, \
  and\ \bibinfo {author} {\bibfnamefont {G.}~\bibnamefont {Kotliar}},\ }\href
  {\doibase 10.1103/PhysRevLett.92.226402} {\bibfield  {journal} {\bibinfo
  {journal} {Physical Review Letters}\ }\textbf {\bibinfo {volume} {92}},\
  \bibinfo {pages} {226402} (\bibinfo {year} {2004})}\BibitemShut {NoStop}%
\bibitem [{\citenamefont {Werner}\ \emph {et~al.}(2009)\citenamefont {Werner},
  \citenamefont {Gull}, \citenamefont {Parcollet},\ and\ \citenamefont
  {Millis}}]{Werner2009}%
  \BibitemOpen
  \bibfield  {author} {\bibinfo {author} {\bibfnamefont {P.}~\bibnamefont
  {Werner}}, \bibinfo {author} {\bibfnamefont {E.}~\bibnamefont {Gull}},
  \bibinfo {author} {\bibfnamefont {O.}~\bibnamefont {Parcollet}}, \ and\
  \bibinfo {author} {\bibfnamefont {A.~J.}\ \bibnamefont {Millis}},\ }\href
  {\doibase 10.1103/PhysRevB.80.045120} {\bibfield  {journal} {\bibinfo
  {journal} {Physical Review B}\ }\textbf {\bibinfo {volume} {80}},\ \bibinfo
  {pages} {045120} (\bibinfo {year} {2009})}\BibitemShut {NoStop}%
\bibitem [{\citenamefont {Biroli}\ \emph {et~al.}(2004)\citenamefont {Biroli},
  \citenamefont {Parcollet},\ and\ \citenamefont {Kotliar}}]{Biroli2004}%
  \BibitemOpen
  \bibfield  {author} {\bibinfo {author} {\bibfnamefont {G.}~\bibnamefont
  {Biroli}}, \bibinfo {author} {\bibfnamefont {O.}~\bibnamefont {Parcollet}}, \
  and\ \bibinfo {author} {\bibfnamefont {G.}~\bibnamefont {Kotliar}},\ }\href
  {\doibase 10.1103/PhysRevB.69.205108} {\bibfield  {journal} {\bibinfo
  {journal} {Physical Review B}\ }\textbf {\bibinfo {volume} {69}},\ \bibinfo
  {pages} {205108} (\bibinfo {year} {2004})}\BibitemShut {NoStop}%
\bibitem [{\citenamefont {Bergeron}\ \emph {et~al.}(2011)\citenamefont
  {Bergeron}, \citenamefont {Hankevych}, \citenamefont {Kyung},\ and\
  \citenamefont {Tremblay}}]{Bergeron2011}%
  \BibitemOpen
  \bibfield  {author} {\bibinfo {author} {\bibfnamefont {D.}~\bibnamefont
  {Bergeron}}, \bibinfo {author} {\bibfnamefont {V.}~\bibnamefont {Hankevych}},
  \bibinfo {author} {\bibfnamefont {B.}~\bibnamefont {Kyung}}, \ and\ \bibinfo
  {author} {\bibfnamefont {A.-M.~S.}\ \bibnamefont {Tremblay}},\ }\href
  {\doibase 10.1103/PhysRevB.84.085128} {\bibfield  {journal} {\bibinfo
  {journal} {Physical Review B}\ }\textbf {\bibinfo {volume} {84}},\ \bibinfo
  {pages} {085128} (\bibinfo {year} {2011})}\BibitemShut {NoStop}%
\bibitem [{\citenamefont {Kyung}\ \emph {et~al.}(2004)\citenamefont {Kyung},
  \citenamefont {Hankevych}, \citenamefont {Dar{\'{e}}},\ and\ \citenamefont
  {Tremblay}}]{Kyung2004}%
  \BibitemOpen
  \bibfield  {author} {\bibinfo {author} {\bibfnamefont {B.}~\bibnamefont
  {Kyung}}, \bibinfo {author} {\bibfnamefont {V.}~\bibnamefont {Hankevych}},
  \bibinfo {author} {\bibfnamefont {A.-M.}\ \bibnamefont {Dar{\'{e}}}}, \ and\
  \bibinfo {author} {\bibfnamefont {A.-M.}\ \bibnamefont {Tremblay}},\ }\href
  {\doibase 10.1103/PhysRevLett.93.147004} {\bibfield  {journal} {\bibinfo
  {journal} {Physical Review Letters}\ }\textbf {\bibinfo {volume} {93}},\
  \bibinfo {pages} {147004} (\bibinfo {year} {2004})}\BibitemShut {NoStop}%
\bibitem [{\citenamefont {Kyung}\ \emph {et~al.}(2006)\citenamefont {Kyung},
  \citenamefont {Kancharla}, \citenamefont {S{\'{e}}n{\'{e}}chal},
  \citenamefont {Tremblay}, \citenamefont {Civelli},\ and\ \citenamefont
  {Kotliar}}]{Kyung2006a}%
  \BibitemOpen
  \bibfield  {author} {\bibinfo {author} {\bibfnamefont {B.}~\bibnamefont
  {Kyung}}, \bibinfo {author} {\bibfnamefont {S.~S.}\ \bibnamefont
  {Kancharla}}, \bibinfo {author} {\bibfnamefont {D.}~\bibnamefont
  {S{\'{e}}n{\'{e}}chal}}, \bibinfo {author} {\bibfnamefont {A.-M.~S.}\
  \bibnamefont {Tremblay}}, \bibinfo {author} {\bibfnamefont {M.}~\bibnamefont
  {Civelli}}, \ and\ \bibinfo {author} {\bibfnamefont {G.}~\bibnamefont
  {Kotliar}},\ }\href {\doibase 10.1103/PhysRevB.73.165114} {\bibfield
  {journal} {\bibinfo  {journal} {Physical Review B}\ }\textbf {\bibinfo
  {volume} {73}},\ \bibinfo {pages} {165114} (\bibinfo {year}
  {2006})}\BibitemShut {NoStop}%
\bibitem [{\citenamefont {Okamoto}\ \emph {et~al.}(2010)\citenamefont
  {Okamoto}, \citenamefont {S{\'{e}}n{\'{e}}chal}, \citenamefont {Civelli},\
  and\ \citenamefont {Tremblay}}]{Okamoto2010}%
  \BibitemOpen
  \bibfield  {author} {\bibinfo {author} {\bibfnamefont {S.}~\bibnamefont
  {Okamoto}}, \bibinfo {author} {\bibfnamefont {D.}~\bibnamefont
  {S{\'{e}}n{\'{e}}chal}}, \bibinfo {author} {\bibfnamefont {M.}~\bibnamefont
  {Civelli}}, \ and\ \bibinfo {author} {\bibfnamefont {A.-M.~S.}\ \bibnamefont
  {Tremblay}},\ }\href {\doibase 10.1103/PhysRevB.82.180511} {\bibfield
  {journal} {\bibinfo  {journal} {Physical Review B}\ }\textbf {\bibinfo
  {volume} {82}},\ \bibinfo {pages} {180511} (\bibinfo {year}
  {2010})}\BibitemShut {NoStop}%
\bibitem [{\citenamefont {Sordi}\ \emph {et~al.}(2010)\citenamefont {Sordi},
  \citenamefont {Haule},\ and\ \citenamefont {Tremblay}}]{Sordi2010}%
  \BibitemOpen
  \bibfield  {author} {\bibinfo {author} {\bibfnamefont {G.}~\bibnamefont
  {Sordi}}, \bibinfo {author} {\bibfnamefont {K.}~\bibnamefont {Haule}}, \ and\
  \bibinfo {author} {\bibfnamefont {A.-M.~S.}\ \bibnamefont {Tremblay}},\
  }\href {\doibase 10.1103/PhysRevLett.104.226402} {\bibfield  {journal}
  {\bibinfo  {journal} {Physical Review Letters}\ }\textbf {\bibinfo {volume}
  {104}},\ \bibinfo {pages} {226402} (\bibinfo {year} {2010})}\BibitemShut
  {NoStop}%
\bibitem [{\citenamefont {Sordi}\ \emph
  {et~al.}(2012{\natexlab{b}})\citenamefont {Sordi}, \citenamefont
  {S{\'{e}}mon}, \citenamefont {Haule},\ and\ \citenamefont
  {Tremblay}}]{Sordi2012a}%
  \BibitemOpen
  \bibfield  {author} {\bibinfo {author} {\bibfnamefont {G.}~\bibnamefont
  {Sordi}}, \bibinfo {author} {\bibfnamefont {P.}~\bibnamefont {S{\'{e}}mon}},
  \bibinfo {author} {\bibfnamefont {K.}~\bibnamefont {Haule}}, \ and\ \bibinfo
  {author} {\bibfnamefont {A.-M.~S.}\ \bibnamefont {Tremblay}},\ }\href
  {\doibase 10.1038/srep00547} {\bibfield  {journal} {\bibinfo  {journal}
  {Scientific Reports}\ }\textbf {\bibinfo {volume} {2}},\ \bibinfo {pages}
  {547} (\bibinfo {year} {2012}{\natexlab{b}})}\BibitemShut {NoStop}%
\bibitem [{\citenamefont {Civelli}\ \emph {et~al.}(2005)\citenamefont
  {Civelli}, \citenamefont {Capone}, \citenamefont {Kancharla}, \citenamefont
  {Parcollet},\ and\ \citenamefont {Kotliar}}]{Civelli2005}%
  \BibitemOpen
  \bibfield  {author} {\bibinfo {author} {\bibfnamefont {M.}~\bibnamefont
  {Civelli}}, \bibinfo {author} {\bibfnamefont {M.}~\bibnamefont {Capone}},
  \bibinfo {author} {\bibfnamefont {S.~S.}\ \bibnamefont {Kancharla}}, \bibinfo
  {author} {\bibfnamefont {O.}~\bibnamefont {Parcollet}}, \ and\ \bibinfo
  {author} {\bibfnamefont {G.}~\bibnamefont {Kotliar}},\ }\href {\doibase
  10.1103/PhysRevLett.95.106402} {\bibfield  {journal} {\bibinfo  {journal}
  {Physical Review Letters}\ }\textbf {\bibinfo {volume} {95}},\ \bibinfo
  {pages} {106402} (\bibinfo {year} {2005})}\BibitemShut {NoStop}%
\bibitem [{\citenamefont {Ferrero}\ \emph {et~al.}(2008)\citenamefont
  {Ferrero}, \citenamefont {Cornaglia}, \citenamefont {{De Leo}}, \citenamefont
  {Parcollet}, \citenamefont {Kotliar},\ and\ \citenamefont
  {Georges}}]{Ferrero2008}%
  \BibitemOpen
  \bibfield  {author} {\bibinfo {author} {\bibfnamefont {M.}~\bibnamefont
  {Ferrero}}, \bibinfo {author} {\bibfnamefont {P.~S.}\ \bibnamefont
  {Cornaglia}}, \bibinfo {author} {\bibfnamefont {L.}~\bibnamefont {{De Leo}}},
  \bibinfo {author} {\bibfnamefont {O.}~\bibnamefont {Parcollet}}, \bibinfo
  {author} {\bibfnamefont {G.}~\bibnamefont {Kotliar}}, \ and\ \bibinfo
  {author} {\bibfnamefont {A.}~\bibnamefont {Georges}},\ }\href {\doibase
  10.1209/0295-5075/85/57009} {\bibfield  {journal} {\bibinfo  {journal}
  {Europhysics Letters}\ }\textbf {\bibinfo {volume} {85}},\ \bibinfo {pages}
  {57009} (\bibinfo {year} {2008})}\BibitemShut {NoStop}%
\bibitem [{\citenamefont {Ferrero}\ \emph {et~al.}(2009)\citenamefont
  {Ferrero}, \citenamefont {Cornaglia}, \citenamefont {{De Leo}}, \citenamefont
  {Parcollet}, \citenamefont {Kotliar},\ and\ \citenamefont
  {Georges}}]{Ferrero2009}%
  \BibitemOpen
  \bibfield  {author} {\bibinfo {author} {\bibfnamefont {M.}~\bibnamefont
  {Ferrero}}, \bibinfo {author} {\bibfnamefont {P.}~\bibnamefont {Cornaglia}},
  \bibinfo {author} {\bibfnamefont {L.}~\bibnamefont {{De Leo}}}, \bibinfo
  {author} {\bibfnamefont {O.}~\bibnamefont {Parcollet}}, \bibinfo {author}
  {\bibfnamefont {G.}~\bibnamefont {Kotliar}}, \ and\ \bibinfo {author}
  {\bibfnamefont {A.}~\bibnamefont {Georges}},\ }\href {\doibase
  10.1103/PhysRevB.80.064501} {\bibfield  {journal} {\bibinfo  {journal}
  {Physical Review B}\ }\textbf {\bibinfo {volume} {80}},\ \bibinfo {pages}
  {064501} (\bibinfo {year} {2009})}\BibitemShut {NoStop}%
\bibitem [{\citenamefont {Gull}\ \emph {et~al.}(2009)\citenamefont {Gull},
  \citenamefont {Parcollet}, \citenamefont {Werner},\ and\ \citenamefont
  {Millis}}]{Gull2009}%
  \BibitemOpen
  \bibfield  {author} {\bibinfo {author} {\bibfnamefont {E.}~\bibnamefont
  {Gull}}, \bibinfo {author} {\bibfnamefont {O.}~\bibnamefont {Parcollet}},
  \bibinfo {author} {\bibfnamefont {P.}~\bibnamefont {Werner}}, \ and\ \bibinfo
  {author} {\bibfnamefont {A.~J.}\ \bibnamefont {Millis}},\ }\href {\doibase
  10.1103/PhysRevB.80.245102} {\bibfield  {journal} {\bibinfo  {journal}
  {Physical Review B}\ }\textbf {\bibinfo {volume} {80}},\ \bibinfo {pages}
  {245102} (\bibinfo {year} {2009})}\BibitemShut {NoStop}%
\bibitem [{\citenamefont {Chen}\ \emph {et~al.}(2015)\citenamefont {Chen},
  \citenamefont {LeBlanc},\ and\ \citenamefont {Gull}}]{Chen2015}%
  \BibitemOpen
  \bibfield  {author} {\bibinfo {author} {\bibfnamefont {X.}~\bibnamefont
  {Chen}}, \bibinfo {author} {\bibfnamefont {J.~P.~F.}\ \bibnamefont
  {LeBlanc}}, \ and\ \bibinfo {author} {\bibfnamefont {E.}~\bibnamefont
  {Gull}},\ }\href {\doibase 10.1103/PhysRevLett.115.116402} {\bibfield
  {journal} {\bibinfo  {journal} {Physical Review Letters}\ }\textbf {\bibinfo
  {volume} {115}},\ \bibinfo {pages} {116402} (\bibinfo {year}
  {2015})}\BibitemShut {NoStop}%
\bibitem [{\citenamefont {Chen}\ \emph {et~al.}(2017)\citenamefont {Chen},
  \citenamefont {LeBlanc},\ and\ \citenamefont {Gull}}]{Chen2016}%
  \BibitemOpen
  \bibfield  {author} {\bibinfo {author} {\bibfnamefont {X.}~\bibnamefont
  {Chen}}, \bibinfo {author} {\bibfnamefont {J.~P.~F.}\ \bibnamefont
  {LeBlanc}}, \ and\ \bibinfo {author} {\bibfnamefont {E.}~\bibnamefont
  {Gull}},\ }\href {\doibase 10.1038/ncomms14986} {\bibfield  {journal}
  {\bibinfo  {journal} {Nature Communications}\ }\textbf {\bibinfo {volume}
  {8}},\ \bibinfo {pages} {14986} (\bibinfo {year} {2017})}\BibitemShut
  {NoStop}%
\bibitem [{\citenamefont {LeBlanc}\ \emph {et~al.}(2015)\citenamefont
  {LeBlanc}, \citenamefont {Antipov}, \citenamefont {Becca}, \citenamefont
  {Bulik}, \citenamefont {Chan}, \citenamefont {Chung}, \citenamefont {Deng},
  \citenamefont {Ferrero}, \citenamefont {Henderson}, \citenamefont
  {Jim{\'{e}}nez-Hoyos}, \citenamefont {Kozik}, \citenamefont {Liu},
  \citenamefont {Millis}, \citenamefont {Prokof'ev}, \citenamefont {Qin},
  \citenamefont {Scuseria}, \citenamefont {Shi}, \citenamefont {Svistunov},
  \citenamefont {Tocchio}, \citenamefont {Tupitsyn}, \citenamefont {White},
  \citenamefont {Zhang}, \citenamefont {Zheng}, \citenamefont {Zhu},\ and\
  \citenamefont {Gull}}]{Leblanc2015}%
  \BibitemOpen
  \bibfield  {author} {\bibinfo {author} {\bibfnamefont {J.~P.~F.}\
  \bibnamefont {LeBlanc}}, \bibinfo {author} {\bibfnamefont {A.~E.}\
  \bibnamefont {Antipov}}, \bibinfo {author} {\bibfnamefont {F.}~\bibnamefont
  {Becca}}, \bibinfo {author} {\bibfnamefont {I.~W.}\ \bibnamefont {Bulik}},
  \bibinfo {author} {\bibfnamefont {G.~K.-L.}\ \bibnamefont {Chan}}, \bibinfo
  {author} {\bibfnamefont {C.-M.}\ \bibnamefont {Chung}}, \bibinfo {author}
  {\bibfnamefont {Y.}~\bibnamefont {Deng}}, \bibinfo {author} {\bibfnamefont
  {M.}~\bibnamefont {Ferrero}}, \bibinfo {author} {\bibfnamefont {T.~M.}\
  \bibnamefont {Henderson}}, \bibinfo {author} {\bibfnamefont {C.~A.}\
  \bibnamefont {Jim{\'{e}}nez-Hoyos}}, \bibinfo {author} {\bibfnamefont
  {E.}~\bibnamefont {Kozik}}, \bibinfo {author} {\bibfnamefont {X.-W.}\
  \bibnamefont {Liu}}, \bibinfo {author} {\bibfnamefont {A.~J.}\ \bibnamefont
  {Millis}}, \bibinfo {author} {\bibfnamefont {N.~V.}\ \bibnamefont
  {Prokof'ev}}, \bibinfo {author} {\bibfnamefont {M.}~\bibnamefont {Qin}},
  \bibinfo {author} {\bibfnamefont {G.~E.}\ \bibnamefont {Scuseria}}, \bibinfo
  {author} {\bibfnamefont {H.}~\bibnamefont {Shi}}, \bibinfo {author}
  {\bibfnamefont {B.~V.}\ \bibnamefont {Svistunov}}, \bibinfo {author}
  {\bibfnamefont {L.~F.}\ \bibnamefont {Tocchio}}, \bibinfo {author}
  {\bibfnamefont {I.~S.}\ \bibnamefont {Tupitsyn}}, \bibinfo {author}
  {\bibfnamefont {S.~R.}\ \bibnamefont {White}}, \bibinfo {author}
  {\bibfnamefont {S.}~\bibnamefont {Zhang}}, \bibinfo {author} {\bibfnamefont
  {B.-X.}\ \bibnamefont {Zheng}}, \bibinfo {author} {\bibfnamefont
  {Z.}~\bibnamefont {Zhu}}, \ and\ \bibinfo {author} {\bibfnamefont
  {E.}~\bibnamefont {Gull}},\ }\href {\doibase 10.1103/PhysRevX.5.041041}
  {\bibfield  {journal} {\bibinfo  {journal} {Physical Review X}\ }\textbf
  {\bibinfo {volume} {5}},\ \bibinfo {pages} {041041} (\bibinfo {year}
  {2015})}\BibitemShut {NoStop}%
\bibitem [{\citenamefont {Gunnarsson}\ \emph {et~al.}(2015)\citenamefont
  {Gunnarsson}, \citenamefont {Sch{\"{a}}fer}, \citenamefont {LeBlanc},
  \citenamefont {Gull}, \citenamefont {Merino}, \citenamefont {Sangiovanni},
  \citenamefont {Rohringer},\ and\ \citenamefont {Toschi}}]{Gunnarsson}%
  \BibitemOpen
  \bibfield  {author} {\bibinfo {author} {\bibfnamefont {O.}~\bibnamefont
  {Gunnarsson}}, \bibinfo {author} {\bibfnamefont {T.}~\bibnamefont
  {Sch{\"{a}}fer}}, \bibinfo {author} {\bibfnamefont {J.~P.~F.}\ \bibnamefont
  {LeBlanc}}, \bibinfo {author} {\bibfnamefont {E.}~\bibnamefont {Gull}},
  \bibinfo {author} {\bibfnamefont {J.}~\bibnamefont {Merino}}, \bibinfo
  {author} {\bibfnamefont {G.}~\bibnamefont {Sangiovanni}}, \bibinfo {author}
  {\bibfnamefont {G.}~\bibnamefont {Rohringer}}, \ and\ \bibinfo {author}
  {\bibfnamefont {A.}~\bibnamefont {Toschi}},\ }\href {\doibase
  10.1103/PhysRevLett.114.236402} {\bibfield  {journal} {\bibinfo  {journal}
  {Physical Review Letters}\ }\textbf {\bibinfo {volume} {114}},\ \bibinfo
  {pages} {236402} (\bibinfo {year} {2015})}\BibitemShut {NoStop}%
\bibitem [{\citenamefont {Gunnarsson}\ \emph {et~al.}()\citenamefont
  {Gunnarsson}, \citenamefont {Sch{\"{a}}fer}, \citenamefont {LeBlanc},
  \citenamefont {Merino}, \citenamefont {Sangiovanni}, \citenamefont
  {Rohringer},\ and\ \citenamefont {Toschi}}]{Gunnarsson2016}%
  \BibitemOpen
  \bibfield  {author} {\bibinfo {author} {\bibfnamefont {O.}~\bibnamefont
  {Gunnarsson}}, \bibinfo {author} {\bibfnamefont {T.}~\bibnamefont
  {Sch{\"{a}}fer}}, \bibinfo {author} {\bibfnamefont {J.~P.~F.}\ \bibnamefont
  {LeBlanc}}, \bibinfo {author} {\bibfnamefont {J.}~\bibnamefont {Merino}},
  \bibinfo {author} {\bibfnamefont {G.}~\bibnamefont {Sangiovanni}}, \bibinfo
  {author} {\bibfnamefont {G.}~\bibnamefont {Rohringer}}, \ and\ \bibinfo
  {author} {\bibfnamefont {A.}~\bibnamefont {Toschi}},\ }\href
  {http://arxiv.org/abs/1604.01614} {\ }\Eprint
  {http://arxiv.org/abs/1604.01614} {arXiv:1604.01614} \BibitemShut {NoStop}%
\bibitem [{\citenamefont {Wu}\ \emph {et~al.}()\citenamefont {Wu},
  \citenamefont {Ferrero}, \citenamefont {Georges},\ and\ \citenamefont
  {Kozik}}]{Wu2016}%
  \BibitemOpen
  \bibfield  {author} {\bibinfo {author} {\bibfnamefont {W.}~\bibnamefont
  {Wu}}, \bibinfo {author} {\bibfnamefont {M.}~\bibnamefont {Ferrero}},
  \bibinfo {author} {\bibfnamefont {A.}~\bibnamefont {Georges}}, \ and\
  \bibinfo {author} {\bibfnamefont {E.}~\bibnamefont {Kozik}},\ }\href
  {http://arxiv.org/abs/1608.08402} {\ }\Eprint
  {http://arxiv.org/abs/1608.08402} {arXiv:1608.08402} \BibitemShut {NoStop}%
\bibitem [{\citenamefont {Rohringer}\ \emph {et~al.}()\citenamefont
  {Rohringer}, \citenamefont {Hafermann}, \citenamefont {Toschi}, \citenamefont
  {Katanin}, \citenamefont {Antipov}, \citenamefont {Katsnelson}, \citenamefont
  {Lichtenstein}, \citenamefont {Rubtsov},\ and\ \citenamefont
  {Held}}]{Rohringer2017}%
  \BibitemOpen
  \bibfield  {author} {\bibinfo {author} {\bibfnamefont {G.}~\bibnamefont
  {Rohringer}}, \bibinfo {author} {\bibfnamefont {H.}~\bibnamefont
  {Hafermann}}, \bibinfo {author} {\bibfnamefont {A.}~\bibnamefont {Toschi}},
  \bibinfo {author} {\bibfnamefont {A.~A.}\ \bibnamefont {Katanin}}, \bibinfo
  {author} {\bibfnamefont {A.~E.}\ \bibnamefont {Antipov}}, \bibinfo {author}
  {\bibfnamefont {M.~I.}\ \bibnamefont {Katsnelson}}, \bibinfo {author}
  {\bibfnamefont {A.~I.}\ \bibnamefont {Lichtenstein}}, \bibinfo {author}
  {\bibfnamefont {A.~N.}\ \bibnamefont {Rubtsov}}, \ and\ \bibinfo {author}
  {\bibfnamefont {K.}~\bibnamefont {Held}},\ }\href
  {http://arxiv.org/abs/1705.00024} {\ }\Eprint
  {http://arxiv.org/abs/1705.00024} {arXiv:1705.00024} \BibitemShut {NoStop}%
\bibitem [{\citenamefont {Biermann}\ \emph {et~al.}(2003)\citenamefont
  {Biermann}, \citenamefont {Aryasetiawan},\ and\ \citenamefont
  {Georges}}]{Biermann2003}%
  \BibitemOpen
  \bibfield  {author} {\bibinfo {author} {\bibfnamefont {S.}~\bibnamefont
  {Biermann}}, \bibinfo {author} {\bibfnamefont {F.}~\bibnamefont
  {Aryasetiawan}}, \ and\ \bibinfo {author} {\bibfnamefont {A.}~\bibnamefont
  {Georges}},\ }\href {\doibase 10.1103/PhysRevLett.90.086402} {\bibfield
  {journal} {\bibinfo  {journal} {Physical Review Letters}\ }\textbf {\bibinfo
  {volume} {90}},\ \bibinfo {pages} {086402} (\bibinfo {year}
  {2003})}\BibitemShut {NoStop}%
\bibitem [{\citenamefont {Sun}\ and\ \citenamefont {Kotliar}(2002)}]{Sun2002}%
  \BibitemOpen
  \bibfield  {author} {\bibinfo {author} {\bibfnamefont {P.}~\bibnamefont
  {Sun}}\ and\ \bibinfo {author} {\bibfnamefont {G.}~\bibnamefont {Kotliar}},\
  }\href {\doibase 10.1103/PhysRevB.66.085120} {\bibfield  {journal} {\bibinfo
  {journal} {Physical Review B}\ }\textbf {\bibinfo {volume} {66}},\ \bibinfo
  {pages} {085120} (\bibinfo {year} {2002})}\BibitemShut {NoStop}%
\bibitem [{\citenamefont {Sun}\ and\ \citenamefont {Kotliar}(2004)}]{Sun2004}%
  \BibitemOpen
  \bibfield  {author} {\bibinfo {author} {\bibfnamefont {P.}~\bibnamefont
  {Sun}}\ and\ \bibinfo {author} {\bibfnamefont {G.}~\bibnamefont {Kotliar}},\
  }\href {http://journals.aps.org/prl/abstract/10.1103/ PhysRevLett.92.196402}
  {\bibfield  {journal} {\bibinfo  {journal} {Physical Review Letters}\
  }\textbf {\bibinfo {volume} {92}},\ \bibinfo {pages} {196402} (\bibinfo
  {year} {2004})}\BibitemShut {NoStop}%
\bibitem [{\citenamefont {Ayral}\ \emph {et~al.}(2012)\citenamefont {Ayral},
  \citenamefont {Werner},\ and\ \citenamefont {Biermann}}]{Ayral2012}%
  \BibitemOpen
  \bibfield  {author} {\bibinfo {author} {\bibfnamefont {T.}~\bibnamefont
  {Ayral}}, \bibinfo {author} {\bibfnamefont {P.}~\bibnamefont {Werner}}, \
  and\ \bibinfo {author} {\bibfnamefont {S.}~\bibnamefont {Biermann}},\ }\href
  {\doibase 10.1103/PhysRevLett.109.226401} {\bibfield  {journal} {\bibinfo
  {journal} {Physical Review Letters}\ }\textbf {\bibinfo {volume} {109}},\
  \bibinfo {pages} {226401} (\bibinfo {year} {2012})}\BibitemShut {NoStop}%
\bibitem [{\citenamefont {Ayral}\ \emph {et~al.}(2013)\citenamefont {Ayral},
  \citenamefont {Biermann},\ and\ \citenamefont {Werner}}]{Ayral2013}%
  \BibitemOpen
  \bibfield  {author} {\bibinfo {author} {\bibfnamefont {T.}~\bibnamefont
  {Ayral}}, \bibinfo {author} {\bibfnamefont {S.}~\bibnamefont {Biermann}}, \
  and\ \bibinfo {author} {\bibfnamefont {P.}~\bibnamefont {Werner}},\ }\href
  {\doibase 10.1103/PhysRevB.87.125149} {\bibfield  {journal} {\bibinfo
  {journal} {Physical Review B}\ }\textbf {\bibinfo {volume} {87}},\ \bibinfo
  {pages} {125149} (\bibinfo {year} {2013})}\BibitemShut {NoStop}%
\bibitem [{\citenamefont {Biermann}(2014)}]{Biermann2014}%
  \BibitemOpen
  \bibfield  {author} {\bibinfo {author} {\bibfnamefont {S.}~\bibnamefont
  {Biermann}},\ }\href {\doibase 10.1088/0953-8984/26/17/173202} {\bibfield
  {journal} {\bibinfo  {journal} {Journal of physics. Condensed matter : an
  Institute of Physics journal}\ }\textbf {\bibinfo {volume} {26}},\ \bibinfo
  {pages} {173202} (\bibinfo {year} {2014})}\BibitemShut {NoStop}%
\bibitem [{\citenamefont {Ayral}\ \emph {et~al.}()\citenamefont {Ayral},
  \citenamefont {Biermann}, \citenamefont {Werner},\ and\ \citenamefont
  {Boehnke}}]{Ayral2017}%
  \BibitemOpen
  \bibfield  {author} {\bibinfo {author} {\bibfnamefont {T.}~\bibnamefont
  {Ayral}}, \bibinfo {author} {\bibfnamefont {S.}~\bibnamefont {Biermann}},
  \bibinfo {author} {\bibfnamefont {P.}~\bibnamefont {Werner}}, \ and\ \bibinfo
  {author} {\bibfnamefont {L.~V.}\ \bibnamefont {Boehnke}},\ }\href
  {https://arxiv.org/abs/1701.07718} {\ }\Eprint
  {http://arxiv.org/abs/1701.07718} {arXiv:1701.07718} \BibitemShut {NoStop}%
\bibitem [{\citenamefont {Rubtsov}\ \emph {et~al.}(2008)\citenamefont
  {Rubtsov}, \citenamefont {Katsnelson},\ and\ \citenamefont
  {Lichtenstein}}]{Rubtsov2008}%
  \BibitemOpen
  \bibfield  {author} {\bibinfo {author} {\bibfnamefont {A.~N.}\ \bibnamefont
  {Rubtsov}}, \bibinfo {author} {\bibfnamefont {M.~I.}\ \bibnamefont
  {Katsnelson}}, \ and\ \bibinfo {author} {\bibfnamefont {A.~I.}\ \bibnamefont
  {Lichtenstein}},\ }\href {\doibase 10.1103/PhysRevB.77.033101} {\bibfield
  {journal} {\bibinfo  {journal} {Physical Review B}\ }\textbf {\bibinfo
  {volume} {77}},\ \bibinfo {pages} {033101} (\bibinfo {year}
  {2008})}\BibitemShut {NoStop}%
\bibitem [{\citenamefont {Rubtsov}\ \emph {et~al.}(2012)\citenamefont
  {Rubtsov}, \citenamefont {Katsnelson},\ and\ \citenamefont
  {Lichtenstein}}]{Rubtsov2011}%
  \BibitemOpen
  \bibfield  {author} {\bibinfo {author} {\bibfnamefont {A.~N.}\ \bibnamefont
  {Rubtsov}}, \bibinfo {author} {\bibfnamefont {M.~I.}\ \bibnamefont
  {Katsnelson}}, \ and\ \bibinfo {author} {\bibfnamefont {A.~I.}\ \bibnamefont
  {Lichtenstein}},\ }\href {http://arxiv.org/abs/1105.6158
  http://www.sciencedirect.com/ science/article/pii/S0003491612000164}
  {\bibfield  {journal} {\bibinfo  {journal} {Annals of Physics}\ }\textbf
  {\bibinfo {volume} {327}},\ \bibinfo {pages} {1320} (\bibinfo {year}
  {2012})}\BibitemShut {NoStop}%
\bibitem [{\citenamefont {van Loon}\ \emph {et~al.}(2014)\citenamefont {van
  Loon}, \citenamefont {Lichtenstein}, \citenamefont {Katsnelson},
  \citenamefont {Parcollet},\ and\ \citenamefont {Hafermann}}]{VanLoon2014}%
  \BibitemOpen
  \bibfield  {author} {\bibinfo {author} {\bibfnamefont {E.~G. C.~P.}\
  \bibnamefont {van Loon}}, \bibinfo {author} {\bibfnamefont {A.~I.}\
  \bibnamefont {Lichtenstein}}, \bibinfo {author} {\bibfnamefont {M.~I.}\
  \bibnamefont {Katsnelson}}, \bibinfo {author} {\bibfnamefont
  {O.}~\bibnamefont {Parcollet}}, \ and\ \bibinfo {author} {\bibfnamefont
  {H.}~\bibnamefont {Hafermann}},\ }\href {\doibase 10.1103/PhysRevB.90.235135}
  {\bibfield  {journal} {\bibinfo  {journal} {Physical Review B}\ }\textbf
  {\bibinfo {volume} {90}},\ \bibinfo {pages} {235135} (\bibinfo {year}
  {2014})}\BibitemShut {NoStop}%
\bibitem [{\citenamefont {Stepanov}\ \emph {et~al.}(2016)\citenamefont
  {Stepanov}, \citenamefont {van Loon}, \citenamefont {Katanin}, \citenamefont
  {Lichtenstein}, \citenamefont {Katsnelson},\ and\ \citenamefont
  {Rubtsov}}]{Stepanov2015}%
  \BibitemOpen
  \bibfield  {author} {\bibinfo {author} {\bibfnamefont {E.~A.}\ \bibnamefont
  {Stepanov}}, \bibinfo {author} {\bibfnamefont {E.~G. C.~P.}\ \bibnamefont
  {van Loon}}, \bibinfo {author} {\bibfnamefont {A.~A.}\ \bibnamefont
  {Katanin}}, \bibinfo {author} {\bibfnamefont {A.~I.}\ \bibnamefont
  {Lichtenstein}}, \bibinfo {author} {\bibfnamefont {M.~I.}\ \bibnamefont
  {Katsnelson}}, \ and\ \bibinfo {author} {\bibfnamefont {A.~N.}\ \bibnamefont
  {Rubtsov}},\ }\href {\doibase 10.1103/PhysRevB.93.045107} {\bibfield
  {journal} {\bibinfo  {journal} {Physical Review B}\ }\textbf {\bibinfo
  {volume} {93}},\ \bibinfo {pages} {045107} (\bibinfo {year}
  {2016})}\BibitemShut {NoStop}%
\bibitem [{\citenamefont {Toschi}\ \emph {et~al.}(2007)\citenamefont {Toschi},
  \citenamefont {Katanin},\ and\ \citenamefont {Held}}]{Toschi2007}%
  \BibitemOpen
  \bibfield  {author} {\bibinfo {author} {\bibfnamefont {A.}~\bibnamefont
  {Toschi}}, \bibinfo {author} {\bibfnamefont {A.}~\bibnamefont {Katanin}}, \
  and\ \bibinfo {author} {\bibfnamefont {K.}~\bibnamefont {Held}},\ }\href
  {\doibase 10.1103/PhysRevB.75.045118} {\bibfield  {journal} {\bibinfo
  {journal} {Physical Review B}\ }\textbf {\bibinfo {volume} {75}},\ \bibinfo
  {pages} {045118} (\bibinfo {year} {2007})}\BibitemShut {NoStop}%
\bibitem [{\citenamefont {Katanin}\ \emph {et~al.}(2009)\citenamefont
  {Katanin}, \citenamefont {Toschi},\ and\ \citenamefont {Held}}]{Katanin2009}%
  \BibitemOpen
  \bibfield  {author} {\bibinfo {author} {\bibfnamefont {A.}~\bibnamefont
  {Katanin}}, \bibinfo {author} {\bibfnamefont {A.}~\bibnamefont {Toschi}}, \
  and\ \bibinfo {author} {\bibfnamefont {K.}~\bibnamefont {Held}},\ }\href
  {\doibase 10.1103/PhysRevB.80.075104} {\bibfield  {journal} {\bibinfo
  {journal} {Physical Review B}\ }\textbf {\bibinfo {volume} {80}},\ \bibinfo
  {pages} {075104} (\bibinfo {year} {2009})}\BibitemShut {NoStop}%
\bibitem [{\citenamefont {Sch{\"{a}}fer}\ \emph {et~al.}(2015)\citenamefont
  {Sch{\"{a}}fer}, \citenamefont {Geles}, \citenamefont {Rost}, \citenamefont
  {Rohringer}, \citenamefont {Arrigoni}, \citenamefont {Held}, \citenamefont
  {Bl{\"{u}}mer}, \citenamefont {Aichhorn},\ and\ \citenamefont
  {Toschi}}]{Schafer2014}%
  \BibitemOpen
  \bibfield  {author} {\bibinfo {author} {\bibfnamefont {T.}~\bibnamefont
  {Sch{\"{a}}fer}}, \bibinfo {author} {\bibfnamefont {F.}~\bibnamefont
  {Geles}}, \bibinfo {author} {\bibfnamefont {D.}~\bibnamefont {Rost}},
  \bibinfo {author} {\bibfnamefont {G.}~\bibnamefont {Rohringer}}, \bibinfo
  {author} {\bibfnamefont {E.}~\bibnamefont {Arrigoni}}, \bibinfo {author}
  {\bibfnamefont {K.}~\bibnamefont {Held}}, \bibinfo {author} {\bibfnamefont
  {N.}~\bibnamefont {Bl{\"{u}}mer}}, \bibinfo {author} {\bibfnamefont
  {M.}~\bibnamefont {Aichhorn}}, \ and\ \bibinfo {author} {\bibfnamefont
  {A.}~\bibnamefont {Toschi}},\ }\href {\doibase 10.1103/PhysRevB.91.125109}
  {\bibfield  {journal} {\bibinfo  {journal} {Physical Review B}\ }\textbf
  {\bibinfo {volume} {91}},\ \bibinfo {pages} {125109} (\bibinfo {year}
  {2015})}\BibitemShut {NoStop}%
\bibitem [{\citenamefont {Valli}\ \emph {et~al.}(2015)\citenamefont {Valli},
  \citenamefont {Sch{\"{a}}fer}, \citenamefont {Thunstr{\"{o}}m}, \citenamefont
  {Rohringer}, \citenamefont {Andergassen}, \citenamefont {Sangiovanni},
  \citenamefont {Held},\ and\ \citenamefont {Toschi}}]{Valli2014}%
  \BibitemOpen
  \bibfield  {author} {\bibinfo {author} {\bibfnamefont {A.}~\bibnamefont
  {Valli}}, \bibinfo {author} {\bibfnamefont {T.}~\bibnamefont
  {Sch{\"{a}}fer}}, \bibinfo {author} {\bibfnamefont {P.}~\bibnamefont
  {Thunstr{\"{o}}m}}, \bibinfo {author} {\bibfnamefont {G.}~\bibnamefont
  {Rohringer}}, \bibinfo {author} {\bibfnamefont {S.}~\bibnamefont
  {Andergassen}}, \bibinfo {author} {\bibfnamefont {G.}~\bibnamefont
  {Sangiovanni}}, \bibinfo {author} {\bibfnamefont {K.}~\bibnamefont {Held}}, \
  and\ \bibinfo {author} {\bibfnamefont {A.}~\bibnamefont {Toschi}},\ }\href
  {\doibase 10.1103/PhysRevB.91.115115} {\bibfield  {journal} {\bibinfo
  {journal} {Physical Review B}\ }\textbf {\bibinfo {volume} {91}},\ \bibinfo
  {pages} {115115} (\bibinfo {year} {2015})}\BibitemShut {NoStop}%
\bibitem [{\citenamefont {Li}\ \emph {et~al.}(2016)\citenamefont {Li},
  \citenamefont {Wentzell}, \citenamefont {Pudleiner}, \citenamefont
  {Thunstr{\"{o}}m},\ and\ \citenamefont {Held}}]{Li2015a}%
  \BibitemOpen
  \bibfield  {author} {\bibinfo {author} {\bibfnamefont {G.}~\bibnamefont
  {Li}}, \bibinfo {author} {\bibfnamefont {N.}~\bibnamefont {Wentzell}},
  \bibinfo {author} {\bibfnamefont {P.}~\bibnamefont {Pudleiner}}, \bibinfo
  {author} {\bibfnamefont {P.}~\bibnamefont {Thunstr{\"{o}}m}}, \ and\ \bibinfo
  {author} {\bibfnamefont {K.}~\bibnamefont {Held}},\ }\href {\doibase
  10.1103/PhysRevB.93.165103} {\bibfield  {journal} {\bibinfo  {journal}
  {Physical Review B}\ }\textbf {\bibinfo {volume} {93}},\ \bibinfo {pages}
  {165103} (\bibinfo {year} {2016})}\BibitemShut {NoStop}%
\bibitem [{\citenamefont {Rohringer}\ and\ \citenamefont
  {Toschi}(2016)}]{Rohringer2016}%
  \BibitemOpen
  \bibfield  {author} {\bibinfo {author} {\bibfnamefont {G.}~\bibnamefont
  {Rohringer}}\ and\ \bibinfo {author} {\bibfnamefont {A.}~\bibnamefont
  {Toschi}},\ }\href {\doibase 10.1103/PhysRevB.94.125144} {\bibfield
  {journal} {\bibinfo  {journal} {Physical Review B}\ }\textbf {\bibinfo
  {volume} {94}},\ \bibinfo {pages} {125144} (\bibinfo {year}
  {2016})}\BibitemShut {NoStop}%
\bibitem [{\citenamefont {Ayral}\ and\ \citenamefont
  {Parcollet}(2016{\natexlab{a}})}]{Ayral2016}%
  \BibitemOpen
  \bibfield  {author} {\bibinfo {author} {\bibfnamefont {T.}~\bibnamefont
  {Ayral}}\ and\ \bibinfo {author} {\bibfnamefont {O.}~\bibnamefont
  {Parcollet}},\ }\href {\doibase 10.1103/PhysRevB.94.075159} {\bibfield
  {journal} {\bibinfo  {journal} {Physical Review B}\ }\textbf {\bibinfo
  {volume} {94}},\ \bibinfo {pages} {075159} (\bibinfo {year}
  {2016}{\natexlab{a}})}\BibitemShut {NoStop}%
\bibitem [{\citenamefont {Hafermann}\ \emph {et~al.}(2008)\citenamefont
  {Hafermann}, \citenamefont {Brener}, \citenamefont {Rubtsov}, \citenamefont
  {Katsnelson},\ and\ \citenamefont {Lichtenstein}}]{Hafermann2008}%
  \BibitemOpen
  \bibfield  {author} {\bibinfo {author} {\bibfnamefont {H.}~\bibnamefont
  {Hafermann}}, \bibinfo {author} {\bibfnamefont {S.}~\bibnamefont {Brener}},
  \bibinfo {author} {\bibfnamefont {A.~N.}\ \bibnamefont {Rubtsov}}, \bibinfo
  {author} {\bibfnamefont {M.~I.}\ \bibnamefont {Katsnelson}}, \ and\ \bibinfo
  {author} {\bibfnamefont {A.~I.}\ \bibnamefont {Lichtenstein}},\ }\href
  {\doibase 10.1134/S0021364007220134} {\bibfield  {journal} {\bibinfo
  {journal} {JETP Letters}\ }\textbf {\bibinfo {volume} {86}},\ \bibinfo
  {pages} {677} (\bibinfo {year} {2008})}\BibitemShut {NoStop}%
\bibitem [{\citenamefont {Slezak}\ \emph {et~al.}(2009)\citenamefont {Slezak},
  \citenamefont {Jarrell}, \citenamefont {Maier},\ and\ \citenamefont
  {Deisz}}]{Slezak2009}%
  \BibitemOpen
  \bibfield  {author} {\bibinfo {author} {\bibfnamefont {C.}~\bibnamefont
  {Slezak}}, \bibinfo {author} {\bibfnamefont {M.}~\bibnamefont {Jarrell}},
  \bibinfo {author} {\bibfnamefont {T.}~\bibnamefont {Maier}}, \ and\ \bibinfo
  {author} {\bibfnamefont {J.}~\bibnamefont {Deisz}},\ }\href {\doibase
  10.1088/0953-8984/21/43/435604} {\bibfield  {journal} {\bibinfo  {journal}
  {Journal of Physics: Condensed Matter}\ }\textbf {\bibinfo {volume} {21}},\
  \bibinfo {pages} {435604} (\bibinfo {year} {2009})}\BibitemShut {NoStop}%
\bibitem [{\citenamefont {Yang}\ \emph
  {et~al.}(2011{\natexlab{b}})\citenamefont {Yang}, \citenamefont {Fotso},
  \citenamefont {Hafermann}, \citenamefont {Tam}, \citenamefont {Moreno},
  \citenamefont {Pruschke},\ and\ \citenamefont {Jarrell}}]{Yang2011c}%
  \BibitemOpen
  \bibfield  {author} {\bibinfo {author} {\bibfnamefont {S.~X.}\ \bibnamefont
  {Yang}}, \bibinfo {author} {\bibfnamefont {H.}~\bibnamefont {Fotso}},
  \bibinfo {author} {\bibfnamefont {H.}~\bibnamefont {Hafermann}}, \bibinfo
  {author} {\bibfnamefont {K.~M.}\ \bibnamefont {Tam}}, \bibinfo {author}
  {\bibfnamefont {J.}~\bibnamefont {Moreno}}, \bibinfo {author} {\bibfnamefont
  {T.}~\bibnamefont {Pruschke}}, \ and\ \bibinfo {author} {\bibfnamefont
  {M.}~\bibnamefont {Jarrell}},\ }\href {\doibase 10.1103/PhysRevB.84.155106}
  {\bibfield  {journal} {\bibinfo  {journal} {Physical Review B}\ }\textbf
  {\bibinfo {volume} {84}},\ \bibinfo {pages} {155106} (\bibinfo {year}
  {2011}{\natexlab{b}})}\BibitemShut {NoStop}%
\bibitem [{\citenamefont {Ayral}\ and\ \citenamefont
  {Parcollet}(2015)}]{Ayral2015}%
  \BibitemOpen
  \bibfield  {author} {\bibinfo {author} {\bibfnamefont {T.}~\bibnamefont
  {Ayral}}\ and\ \bibinfo {author} {\bibfnamefont {O.}~\bibnamefont
  {Parcollet}},\ }\href {\doibase 10.1103/PhysRevB.92.115109} {\bibfield
  {journal} {\bibinfo  {journal} {Physical Review B}\ }\textbf {\bibinfo
  {volume} {92}},\ \bibinfo {pages} {115109} (\bibinfo {year}
  {2015})}\BibitemShut {NoStop}%
\bibitem [{\citenamefont {Ayral}\ and\ \citenamefont
  {Parcollet}(2016{\natexlab{b}})}]{Ayral2015c}%
  \BibitemOpen
  \bibfield  {author} {\bibinfo {author} {\bibfnamefont {T.}~\bibnamefont
  {Ayral}}\ and\ \bibinfo {author} {\bibfnamefont {O.}~\bibnamefont
  {Parcollet}},\ }\href {\doibase 10.1103/PhysRevB.93.235124} {\bibfield
  {journal} {\bibinfo  {journal} {Physical Review B}\ }\textbf {\bibinfo
  {volume} {93}},\ \bibinfo {pages} {235124} (\bibinfo {year}
  {2016}{\natexlab{b}})}\BibitemShut {NoStop}%
\bibitem [{\citenamefont {Vu{\v{c}}i{\v{c}}evi{\'{c}}}\ \emph
  {et~al.}()\citenamefont {Vu{\v{c}}i{\v{c}}evi{\'{c}}}, \citenamefont
  {Ayral},\ and\ \citenamefont {Parcollet}}]{Vucicevic2017}%
  \BibitemOpen
  \bibfield  {author} {\bibinfo {author} {\bibfnamefont {J.}~\bibnamefont
  {Vu{\v{c}}i{\v{c}}evi{\'{c}}}}, \bibinfo {author} {\bibfnamefont
  {T.}~\bibnamefont {Ayral}}, \ and\ \bibinfo {author} {\bibfnamefont
  {O.}~\bibnamefont {Parcollet}},\ }\href {https://arxiv.org/abs/1705.08332} {\
  }\Eprint {http://arxiv.org/abs/1705.08332} {arXiv:1705.08332} \BibitemShut
  {NoStop}%
\bibitem [{Note1()}]{Note1}%
  \BibitemOpen
  \bibinfo {note} {See Suppl. Mat. I for another choice}\BibitemShut {NoStop}%
\bibitem [{\citenamefont {Park}\ \emph {et~al.}(2008)\citenamefont {Park},
  \citenamefont {Haule},\ and\ \citenamefont {Kotliar}}]{Park2008}%
  \BibitemOpen
  \bibfield  {author} {\bibinfo {author} {\bibfnamefont {H.}~\bibnamefont
  {Park}}, \bibinfo {author} {\bibfnamefont {K.}~\bibnamefont {Haule}}, \ and\
  \bibinfo {author} {\bibfnamefont {G.}~\bibnamefont {Kotliar}},\ }\href
  {\doibase 10.1103/PhysRevLett.101.186403} {\bibfield  {journal} {\bibinfo
  {journal} {Physical Review Letters}\ }\textbf {\bibinfo {volume} {101}},\
  \bibinfo {pages} {186403} (\bibinfo {year} {2008})}\BibitemShut {NoStop}%
\bibitem [{\citenamefont {Blankenbecler}\ \emph {et~al.}(1981)\citenamefont
  {Blankenbecler}, \citenamefont {Scalapino},\ and\ \citenamefont
  {Sugar}}]{Blankenbecler1981}%
  \BibitemOpen
  \bibfield  {author} {\bibinfo {author} {\bibfnamefont {R.}~\bibnamefont
  {Blankenbecler}}, \bibinfo {author} {\bibfnamefont {D.~J.}\ \bibnamefont
  {Scalapino}}, \ and\ \bibinfo {author} {\bibfnamefont {R.~L.}\ \bibnamefont
  {Sugar}},\ }\href {\doibase 10.1103/PhysRevD.24.2278} {\bibfield  {journal}
  {\bibinfo  {journal} {Physical Review D}\ }\textbf {\bibinfo {volume} {24}},\
  \bibinfo {pages} {2278} (\bibinfo {year} {1981})}\BibitemShut {NoStop}%
\bibitem [{Note2()}]{Note2}%
  \BibitemOpen
  \bibinfo {note} {This also holds for other HS decouplings, Suppl. Mat.
  I}\BibitemShut {NoStop}%
\bibitem [{\citenamefont {Parcollet}\ \emph {et~al.}(2015)\citenamefont
  {Parcollet}, \citenamefont {Ferrero}, \citenamefont {Ayral}, \citenamefont
  {Hafermann}, \citenamefont {Seth},\ and\ \citenamefont
  {Krivenko}}]{Parcollet2014}%
  \BibitemOpen
  \bibfield  {author} {\bibinfo {author} {\bibfnamefont {O.}~\bibnamefont
  {Parcollet}}, \bibinfo {author} {\bibfnamefont {M.}~\bibnamefont {Ferrero}},
  \bibinfo {author} {\bibfnamefont {T.}~\bibnamefont {Ayral}}, \bibinfo
  {author} {\bibfnamefont {H.}~\bibnamefont {Hafermann}}, \bibinfo {author}
  {\bibfnamefont {P.}~\bibnamefont {Seth}}, \ and\ \bibinfo {author}
  {\bibfnamefont {I.~S.}\ \bibnamefont {Krivenko}},\ }\href {\doibase
  10.1016/j.cpc.2015.04.023} {\bibfield  {journal} {\bibinfo  {journal}
  {Computer Physics Communications}\ }\textbf {\bibinfo {volume} {196}},\
  \bibinfo {pages} {398} (\bibinfo {year} {2015})}\BibitemShut {NoStop}%
\bibitem [{\citenamefont {Rubtsov}\ \emph {et~al.}(2005)\citenamefont
  {Rubtsov}, \citenamefont {Savkin},\ and\ \citenamefont
  {Lichtenstein}}]{Rubtsov2005}%
  \BibitemOpen
  \bibfield  {author} {\bibinfo {author} {\bibfnamefont {A.~N.}\ \bibnamefont
  {Rubtsov}}, \bibinfo {author} {\bibfnamefont {V.~V.}\ \bibnamefont {Savkin}},
  \ and\ \bibinfo {author} {\bibfnamefont {A.~I.}\ \bibnamefont
  {Lichtenstein}},\ }\href {\doibase 10.1103/PhysRevB.72.035122} {\bibfield
  {journal} {\bibinfo  {journal} {Physical Review B}\ }\textbf {\bibinfo
  {volume} {72}},\ \bibinfo {pages} {035122} (\bibinfo {year}
  {2005})}\BibitemShut {NoStop}%
\bibitem [{\citenamefont {Gull}\ \emph {et~al.}(2011)\citenamefont {Gull},
  \citenamefont {Millis}, \citenamefont {Lichtenstein}, \citenamefont
  {Rubtsov}, \citenamefont {Troyer},\ and\ \citenamefont {Werner}}]{Gull2011}%
  \BibitemOpen
  \bibfield  {author} {\bibinfo {author} {\bibfnamefont {E.}~\bibnamefont
  {Gull}}, \bibinfo {author} {\bibfnamefont {A.~J.}\ \bibnamefont {Millis}},
  \bibinfo {author} {\bibfnamefont {A.~I.}\ \bibnamefont {Lichtenstein}},
  \bibinfo {author} {\bibfnamefont {A.~N.}\ \bibnamefont {Rubtsov}}, \bibinfo
  {author} {\bibfnamefont {M.}~\bibnamefont {Troyer}}, \ and\ \bibinfo {author}
  {\bibfnamefont {P.}~\bibnamefont {Werner}},\ }\href {\doibase
  10.1103/RevModPhys.83.349} {\bibfield  {journal} {\bibinfo  {journal}
  {Reviews of Modern Physics}\ }\textbf {\bibinfo {volume} {83}},\ \bibinfo
  {pages} {349} (\bibinfo {year} {2011})}\BibitemShut {NoStop}%
\bibitem [{\citenamefont {Staar}\ \emph {et~al.}(2013)\citenamefont {Staar},
  \citenamefont {Maier},\ and\ \citenamefont {Schulthess}}]{Staar2013}%
  \BibitemOpen
  \bibfield  {author} {\bibinfo {author} {\bibfnamefont {P.}~\bibnamefont
  {Staar}}, \bibinfo {author} {\bibfnamefont {T.}~\bibnamefont {Maier}}, \ and\
  \bibinfo {author} {\bibfnamefont {T.~C.}\ \bibnamefont {Schulthess}},\ }\href
  {\doibase 10.1103/PhysRevB.88.115101} {\bibfield  {journal} {\bibinfo
  {journal} {Physical Review B}\ }\textbf {\bibinfo {volume} {88}},\ \bibinfo
  {pages} {115101} (\bibinfo {year} {2013})}\BibitemShut {NoStop}%
\bibitem [{\citenamefont {Staar}\ \emph {et~al.}(2014)\citenamefont {Staar},
  \citenamefont {Maier},\ and\ \citenamefont {Schulthess}}]{Staar2014}%
  \BibitemOpen
  \bibfield  {author} {\bibinfo {author} {\bibfnamefont {P.}~\bibnamefont
  {Staar}}, \bibinfo {author} {\bibfnamefont {T.}~\bibnamefont {Maier}}, \ and\
  \bibinfo {author} {\bibfnamefont {T.~C.}\ \bibnamefont {Schulthess}},\ }\href
  {\doibase 10.1103/PhysRevB.89.195133} {\bibfield  {journal} {\bibinfo
  {journal} {Physical Review B}\ }\textbf {\bibinfo {volume} {89}},\ \bibinfo
  {pages} {195133} (\bibinfo {year} {2014})}\BibitemShut {NoStop}%
\end{thebibliography}%

\end{document}